\documentclass[twocolumn]{aastex62}

\hypersetup{linkcolor=blue,citecolor=blue,filecolor=blue,urlcolor=blue}

\graphicspath{{./}{figures/}}

\received{XXX}
\revised{XXX}
\accepted{XXX}
\submitjournal{ApJ}

\shorttitle{Setting the Stage for Cosmic Chronometers II}
\shortauthors{Moresco M. et al.}

\begin{document}
\title{Setting the Stage for Cosmic Chronometers. II.\\ Impact of Stellar Population Synthesis Models Systematics and Full Covariance Matrix}

\author{Michele Moresco}
\affil{Dipartimento di Fisica e Astronomia, Universit\`a di Bologna, Via Gobetti 93/2, I-40129, Bologna, Italy}
\affiliation{INAF - Osservatorio di Astrofisica e Scienza dello Spazio di Bologna, via Gobetti 93/3, I-40129 Bologna, Italy}

\author{Raul Jimenez}
\affiliation{ICC, Instituto de Ciencias del Cosmos, University of Barcelona, UB, Marti i Franques 1, E-08028, Barcelona, Spain}
\affiliation{ICREA, Pg. Lluis Companys 23, E-08010 Barcelona, Spain}

\author{Licia Verde}
\affiliation{ICC, Instituto de Ciencias del Cosmos, University of Barcelona, UB, Marti i Franques 1, E-08028, Barcelona, Spain}
\affiliation{ICREA, Pg. Lluis Companys 23, E-08010 Barcelona, Spain}

\author{Andrea Cimatti}
\affiliation{Dipartimento di Fisica e Astronomia, Universit\`a di Bologna, Via Gobetti 93/2, I-40129, Bologna, Italy}
\affiliation{INAF - Osservatorio Astrofisico di Arcetri, Largo E. Fermi 5, I-50125 Firenze, Italy}

\author{Lucia Pozzetti}
\affiliation{INAF - Osservatorio di Astrofisica e Scienza dello Spazio di Bologna, via Gobetti 93/3, I-40129 Bologna, Italy}

\correspondingauthor{Michele Moresco}
\email{michele.moresco@unibo.it}



\begin{abstract}
The evolution of differential ages of passive galaxies at different redshifts (cosmic chronometers) has been proved to be a method potentially able to constrain the Hubble parameter in a cosmology-independent way, but the systematic uncertainties must be carefully evaluated. 
In this paper, we compute the contribution to the full covariance matrix of systematic uncertainties due to the choice of initial mass function, stellar library, and metallicity, exploring a variety of stellar population synthesis models. 
Through simulations in the redshift range $0<z<1.5$ we find that the choice of the stellar population synthesis model dominates the total error budget on $H(z)$, with contributions at a level of $\sim$4.5\%, discarding the most discordant model. The contribution due to the choice of initial mass function is $<$0.5\%, while that due to the stellar library is $\sim$6.6\% on average. 
We also assess the impact of an uncertainty in the stellar metallicity determination, finding that an error of $\sim$10\% (5\%) on the stellar metallicity propagates to a 9\% (4\%) error on $H(z)$. These results are used to provide the combined contribution of these systematic effects on the error budget. For current $H(z)$ measurements, where the uncertainties due to metallicity and star formation history were already included, we show that, using the more modern stellar libraries, the additional systematic uncertainty is between 5.4\% (at $z=0.2$) and 2.3\% (at $z=1.5$). 
To reach the goal of keeping the systematic error budget below the 1\% level we discuss the efforts needed to obtain higher resolution and signal-to-noise spectra and improvements in the modeling of stellar population synthesis.	
\end{abstract}


\section{Introduction}

The cosmic chronometers (CC) method is a conceptually simple technique to measure the Hubble parameter as a function of redshift, $H(z)$, independent of the cosmological model adopted \citep{jimenez2002}. The method is based on the relationship between time and redshift, which for a Friedmann-Robertson-Walker metric is
\begin{equation}
H(z) = -\frac{1}{(1+z)} \frac{dz}{dt}
\label{eq:Hz}
\end{equation}
providing a route to measure $H(z)$ in a cosmology-independent way, if $dt$ and $dz$ can be obtained with sufficient precision. While the redshift can be measured to an accuracy $\delta z/z \lesssim 0.001$ with spectroscopy of extragalactic objects, the main difficulty is to obtain a robust estimate of the differential age evolution $dt$. This requires the use of a ``chronometer''. 
The ideal candidates to be exploited as CCs are passive stellar populations that are evolving on a much longer time-scale compared to their age difference.
Massive (${\rm log(M/M_{\odot})\gtrsim11}$) and passive early-type galaxies represent, therefore, the best option, since many independent analyses have found that typically, they have formed and assembled their mass at high redshifts ($z>2-3$) and over a relatively short period of time ($\lesssim$0.3 Gyr), and that, having exhausted their gas reservoir in the early stages of their life, they are mostly passively evolving \citep{cimatti2004,treu2005,pozzetti2010,thomas2010,choi2014,onodera2015,citro2016,pacifici2016,belli2019,carnall2019,estrada2019}. As a consequence, they constitute the oldest population of galaxies at each redshift and can be therefore used to homogeneously trace the differential age evolution of the universe ($dt$) as a function of redshift \citep[for an extensive review, see][and references therein]{renzini2006}.

In previous works \citep{simon2005,carson2010,stern2010,liu2012,moresco2012,zhang2014,moresco2015,moresco2016,ratsimbazafy2017}, it was demonstrated how this method can be applied to galaxy surveys over a wide range of redshifts, $0.1 < z < 2$, and that the statistical uncertainty from current galaxy samples can lead to a determination of $H(z)$ at the $5\%$ accuracy level. 

The main strength of the CC approach is that it provides a direct estimate of the expansion history of the universe without relying on any cosmological assumption, providing an ideal framework to test cosmological models.
These results have been extensively used to provide constraints on several cosmological parameters, both in standard and alternative cosmological models and in combination with other standard probes \citep{moresco2012b,seikel2012,capozziello2014,sapone2014,valkenburg2014,nunes2016,lhuillier2017,moresco2017,sola2017,yang2018,lin2019},
to explore a possible time evolution of the dark energy equation-of-state parameter \citep{moresco2016b,zhao2017,yang2018}, and, more recently, also in the context of the Hubble constant controversy \citep{verde2019} to provide an independent constraint on $H_0$ \citep{gomez2018,haridasu2018,jimenez2019}.
Forecasts for future galaxy surveys estimate that $H(z)$ can be recovered at the percent level \citep[see, e.g.,][]{ma2011,moresco2015}.
We emphasize that current surveys are not optimized to obtain spectra for most passively evolving galaxies, especially spectra that allow for highly accurate extraction of the stellar population parameters of the galaxy. 

The critical obstacle in measuring $H(z)$ with the CC method is not the statistical but the systematic uncertainty, which can be divided into four main sources: (i) the one depending on the stellar population synthesis (SPS) model used to calibrate the measurement, (ii) the one depending on the estimate of the stellar metallicity of the population, (iii) the one depending on the assumed star formation history (SFH) of the adopted model, and (iv) the one depending on a possible residual star formation due to a young subdominant component underlying in the sample selected.

An initial assessment of the impact of systematic uncertainties was done in \citet{moresco2012,moresco2016}, where it was quantified the impact of SFH assumption in the method to be between 2\% and 3\% ($2.5$\% on average). In the first article of this series of papers \citep[][hereafter Paper I]{moresco2018} we focused on the impact of a recent burst of star formation (``frosting'') on a carefully selected sample of passively evolving galaxies based on CaII H and K lines. There it is demonstrated that, even with only optical spectra (in the restframe), it is possible to minimize this effect through a careful selection of purely passively evolving galaxies. In Paper I the effect of this systematic was quantified and provided an analytic formula to compute its contribution to the $H(z)$ errors and covariance matrix; the recommended selection procedure limits the frosting-induced systematic error in the selected sample to less than 0.5\% in $H(z)$.

The remaining dominant systematic contributions are due to the choice of the SPS model (i.e., stellar physics models, along with an adopted stellar library, initial mass function (IMF), etc.) and the  metallicity. The goal of this paper is to quantify these effects and to provide a comprehensive assessment of the uncertainties in the CC method.

This article is organized as follows. In Section~\ref{sec:method} we describe the method to compute the covariances; in Section~\ref{sec:results} we present the main results and lessons learned from the computed covariance matrices. In Section~\ref{sec:Hzmeas} we present how all of our results combine to provide a clear estimate of the total covariance for the method, and in Section~\ref{sec:examples} we provide illustrative examples on how to apply this formalism. We conclude in Section~\ref{sec:conclusions}. Throughout this paper, we will assume a fiducial $\Lambda$CDM cosmology from \citet{planck2016}\footnote{Note that the results do not depend on the assumed fiducial cosmology, which is currently only used as a reference.}.

\section{Method}
\label{sec:method}

A way forward in improving the CC method is to find a stable and robust way to estimate the differential ages $dt$ in Equation~\ref{eq:Hz}. An option, suggested firstly in \citet{moresco2011}, is to study a direct observable in galaxy spectra instead of relying on the estimate of the age of the stellar population from a fit. With this approach, it is possible to achieve an easier and more transparent disentanglement of statistical and systematic errors.
In particular, it was shown that the 4000~\AA~break ($D4000$) is a spectral feature that can be adopted as an age indicator \citep{hamilton1985,poggianti1997,balogh1999}. One of the advantages of this feature is that (in given intervals) it correlates almost linearly with the age of the population (at fixed metallicity $Z$), so that, by differentiating this relation, it is possible to rewrite $dt=A(Z, SFH)\times dD4000$, where $A(Z, SFH)$ is the slope of the $D4000-$age relation for the metallicity $Z$ (assuming a given SFH).

Under this assumption \citep[which has been demonstrated to be a good proxy of the actual theoretical trend; see, e.g.,][]{moresco2012,moresco2016}, it is possible to rewrite Equation~\ref{eq:Hz} as
\begin{equation}
H(z) = -\frac{A(Z,SFH)}{(1+z)} \frac{dz}{dD4000}\; .
\label{eq:Hz_D4000}
\end{equation}
It is therefore easy to understand what the dominant systematic errors in the $H(z)$ determination are: frosting from a young component affects the galaxy spectrum and could therefore bias the measurement of $dD$4000; SPS models are used to provide the parameter $A$ as the slope of the $D4000-$age relation, and could, therefore, bias this quantity; and the metallicity determination is used at a fixed SPS model to obtain the parameter $A$ calibrated for the appropriate metallicity. 

\begin{figure*}[t!]
	\centering
	\includegraphics[width=0.75\textwidth]{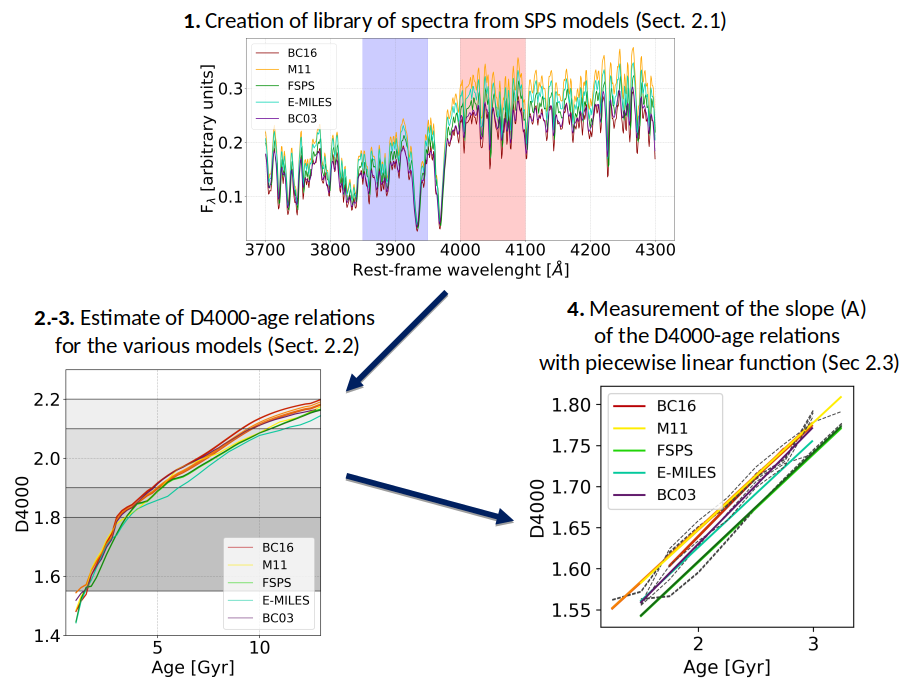}
	\includegraphics[width=0.75\textwidth]{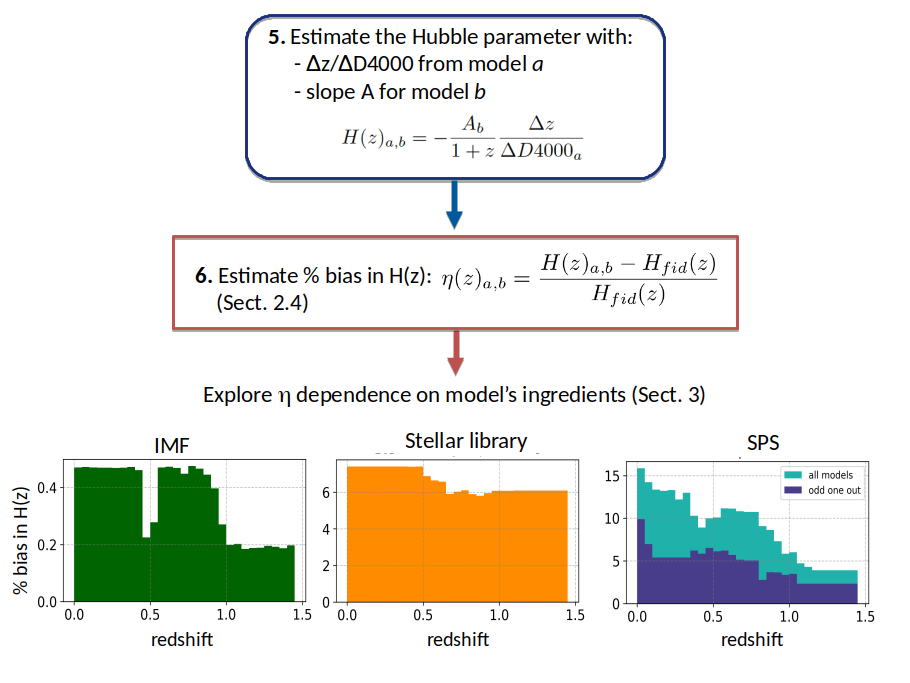}
	\caption{Analysis workflow. This diagram shows the six main steps to assess the impact of systematic uncertainties on the $H(z)$ determination with the CC method. At first, a library of synthetic spectra is generated considering a wide range of SPS models. Then, the $D4000$ of these spectra is measured (for each SPS model) at various ages to construct the $D4000-$age relations. Next, these relations are fitted with a piecewise linear function to obtain the slope $A$ of the relation for each model. These two measurements are then combined in Equation~\ref{eq:Hz_D4000} to obtain and $H(z)_{a,b}$, extracting the $\Delta D4000$ from a model $a$ and the slope $A$ from a model $b$. This measurement is iterated on the various possible combination of models to build the percentage bias matrix. Finally, this matrix is studied as a function of the various models' ingredients to assess the impact of each one on the final $H(z)$ measurement. We also report for each step the corresponding section where it is discussed.}
	\label{fig:workflow}
\end{figure*}

The covariance matrix associated with the CC method can therefore be expressed as
\begin{equation}
{\rm Cov}_{ij}= {\rm Cov}_{ij}^{\rm stat}+ {\rm Cov}_{ij}^{\rm young}+ {\rm Cov}_{ij}^{\rm model}+ {\rm Cov}_{ij}^{\rm met}
\end{equation}
where ``stat,'' ``young,'' ``model,'' and ``met'' denote the contributions to the covariance due to statistical errors, young component contamination, dependence on the chosen model, and stellar metallicity, respectively. The contribution due to model ${\rm Cov}_{ij}^{\rm model}$ can be further decomposed in the contribution due to SFH, IMF, stellar library, and SPS model considered as
\begin{equation}
\label{eq:Covmodel}
{\rm Cov}_{ij}^{\rm model}={\rm Cov}_{ij}^{\rm SFH}+{\rm Cov}_{ij}^{\rm IMF}+{\rm Cov}_{ij}^{\rm st. lib.}+{\rm Cov}_{ij}^{\rm SPS} \; .
\end{equation}
Since, as discussed, the young component contamination and the SFH dependence were already computed in previous papers \citep{moresco2016,moresco2018}, in this analysis, we focus on the other two systematic terms.
In the following, we refer to a choice of SPS model with a given stellar library, metallicity, and IMF as a ``model,'' which is labeled by indices $a$ or $b$.

Given the observed spectra of CCs, to obtain a measurement of $H(z)$ at a given redshift $z$, two ingredients are needed (following Equation~\ref{eq:Hz_D4000}): (1) from the data, an estimate of the differential $D4000$ evolution of CCs between two redshifts, $z_1$ and $z_2$, both close to the redshift $z$ of interest, i.e., $\Delta z/\Delta D$4000; and (2) from SPS models, the slope of the $D4000-$age relation. 
Therefore, to assess the impact of SPS models on the $H(z)$ estimate, we simulate ``mock'' $D4000$ measurements across a range of  redshifts and ages and then fit them assuming different SPS models.

The main steps needed to estimate these last contributions to the covariance can therefore be summarized as follows.
\begin{enumerate}
\item A library of synthetic simple stellar population (SSP) spectra is generated spanning a wide range of properties (SPS models, stellar ages and metallicities, IMFs, stellar libraries; see Section \ref{sec:lib}). In the following, we label each different SSP model with letters, e.g., $a,b$. We note here that in this step, we are considering SSP models, since the dependence of the systematic errors on the SFH has been assessed separately.
\item The $D4000$ is measured for all spectra in the library introduced in step 1 to build the $D4000-$age relation for various models (see Section \ref{sec:measD4000}). 
\item The relations obtained in step 2 are used to generate $\Delta z/\Delta D4000$ measurements from ``mock'' (noiseless) simulations of $D4000(z)$, for each model in the library (see Section \ref{sec:measD4000}).
\item The slopes of the $D4000$-age relations, $A(Z)$,  discussed in step 2 are estimated with a piecewise linear fit (see Section \ref{sec:measslope}) obtained by performing a linear fit in different $D4000$ ranges. This approach provides, for each model, several slopes $A$ as a function of the ranges of $D4000$ in which the piecewise linear fit is performed.
\item The Hubble parameter $H(z)$ is estimated by extracting the differential $\Delta D4000$ from ``mock''  realizations generated with model $a$, and the slope from the model $b$, and the percentage $H(z)$ bias matrix $\eta(z)_{ab}$ is constructed for various model combinations (see Section \ref{sec:measmatrix}).
\item The percentage $H(z)$ bias matrix $\eta(z)_{ab}$ is then propagated on a mean percentage bias $\widehat{\eta}(z)$ (and its correlations) on $H(z)$ as a function of redshift (see Section \ref{sec:results}).
\end{enumerate}
This approach allows us to isolate and quantify the impact of each ingredient of the model (IMF, stellar library, and SPS model) on the final error budget. The covariance matrix ${\rm Cov}_{ij}^{\rm model}$ as a function of each ingredient of the model can therefore be estimated from the mean percentage bias $\widehat{\eta}(z)$.

The general workflow of the analysis is summarized in Figure \ref{fig:workflow}. In the rest of this section, we will discuss these steps separately.

\subsection{Creation of the Library of SPS Model Spectra}
\label{sec:lib}

We consider a variety of SPS models that are usually adopted in galaxy evolution studies, whose properties are summarized in Table~\ref{tab:tab1}.
The library includes the Bruzual \& Charlot updated 2016 models \citep[hereafter BC16, which represents an update of the ][models, hereafter BC03]{bruzual2003}, the \cite{maraston2011} models (hereafter M11), the extended MILES models \citep[hereafter E-MILES;][]{vazdekis2016}, and the Flexible Stellar Population Synthesis models \citep[hereafter FSPS;][]{conroy2009,conroy2010}.
We also include the BC03 models, since many $H(z)$ measurements available in the literature have been obtained from those models. In all cases, we are considering solar-scaled chemical mixtures, since $\alpha$-enhanced chemical mixtures are not available for all models, and in \cite{moresco2012} it was shown that it has a minor impact on the results.

This library encompasses different recipes and physical assumptions, in this way spanning a wide range of possibilities. Below, we will describe the adopted ingredients for each model, but for a detailed comparison, we refer to \cite{baldwin2018}, where an extensive comparison has been done based on optical and infrared spectroscopic data. 

\paragraph{BC16 and BC03 models.} The
Bruzual \& Charlot models are built with the isochrone synthesis technique and use the Padova tracks \citep{alongi1993,bressan1993,fagotto1994}. The updated BC16 models have been generated considering different stellar libraries, and in this work, we considered both the newest MILES and the older STELIB stellar libraries, with spectral resolutions of 2.3 and 3~\AA, respectively \citep[see][]{bruzual2003}. 
The IMFs included in the analysis are the Chabrier, Kroupa and Salpeter one \citep{kroupa2001,chabrier2003,salpeter1955}, and the stellar metallicities adopted are a solar ($Z_\odot\approx0.02$), a subsolar ($Z/Z_{\odot}\approx0.4$) and a supersolar ($Z/Z_{\odot}\approx2.5$).
We include in the library also the BC03 models, considering Padova isochrones, a STELIB stellar library, a Chabrier IMF, and the same grid of stellar metallicities as BC16.

\paragraph{M11 models.}
The M11 models are based on the fuel consumption theorem \citep{renzini1981}, and thermally pulsing (TP) asymptotic giant branch (AGB) phases stars are included based on \cite{lancon2002}. The M11 models have been constructed by using the MILES stellar library, the \citet{cassisi1997} isochrones, and a Chabrier IMF and taking into account three stellar metallicities available close to the solar one, namely $Z/Z_{\odot}=$0.5, 1, and 2.

\paragraph{FSPS models.}
Similar to M11 models, the FSPS models include TP-AGB phases following \cite{lancon2002}, and in addition, they also consider circumstellar dust shells around AGB stars \citep{villaume2015}. To create FSPS models, we consider the MILES stellar library and Padova isochrones, also spanning, in this case, a variety of IMFs (Chabrier, Kroupa, and Salpeter).
The metallicity available by default with the Padova isochrones differs slightly from the ones of BC16 models, and in order not to bias our results by this effect, we interpolate the models at BC16 solar metallicity, taking advantage of the Python version of the FSPS code. We take particular care in trying to calibrate to the same values of solar metallicity (which is the main focus of this analysis), thus sampling the following metallicity grid points: $Z/Z_{\odot}=$0.5, 1, and 1.5.

\paragraph{E-MILES models.}
The E-MILES models adopt the synthetic AGB technique, mapping the AGB stages up to the TP-AGB phase. In this case, we use BaSTI isochrones, the MILES stellar library, a Chabrier IMF, and also, in this case, the closest stellar metallicities available close to the solar values, even if we note here that they slightly differ from the other cases for the solar value ($Z/Z_{\odot}=$0.5, 0.99, and 2).

\begin{table*}[t!]
	\centering
	\caption{Library of SPS Models and Corresponding Characteristics.}
	\begin{tabular}{cccccc}
		\hline
		\hline
		& BC16 & M11 & FSPS & E-MILES & BC03\\
		\hline
		Ages [Gyr] & [1-13] & [1-13] & [1-13] & [1-13] & [1-13]\\
		$Z$ & 0.008, 0.02, 0.05 & 0.01, 0.02, 0.04 & 0.01, 0.02, 0.03 & 0.010, 0.0198, 0.04 & 0.008, 0.02, 0.05\\
		$\rm[Fe/H]$ & -0.330, 0.093, & -0.330, 0., 0.35 & -0.279, 0.022, 0.198 & -0.250, 0.060, 0.400 & -0.330, +0.093,\\ 
		& 0.560 & & & & 0.560\\
		Stellar library & MILES, STELIB & MILES & MILES & E-MILES & STELIB\\
		IMF & Chabrier & Chabrier & Chabrier & Chabrier & Chabrier \\
		& Salpeter, Kroupa & & Salpeter, Kroupa & & \\
		\hline
	\end{tabular}
	\label{tab:tab1}
\end{table*}

To summarize, as Table~\ref{tab:tab1} indicates, we have collected a total of 12 possible combinations of SPS comprised of five SPS models, of which BC16 has the choice of two different stellar libraries (MILES and STELIB), and both BC16 and FPS have the choice of three different IMFs; in particular, the MILES stellar library is in common between BC16, M11, FSPS, and E-MILES. This allows us to quantify the contribution to the total error due to a variation of SPS model, stellar library, and IMF. We refer to the original papers for a more extensive discussion of each model.

For each model, we extract SSP synthetic spectra with ages spanning $\rm1\leq t[Gyr]\leq13$ at the maximum age resolution allowed (0.25 Gyr for all models except for the E-MILES models above 4 Gyr, where the age resolution is 0.5 Gyr). In this analysis, we consider SSP spectra for two reasons: (i) because we provided in other papers the technique to propagate to the final measurement an uncertainty due to this effect \citep{moresco2012,moresco2016}, demonstrating that the uncertainty on the SFH of the CC population impacts the estimate of $H(z)$ at a 2-3\% level; and (ii) because the SFH of these systems is, however, found to be extremely rapid and focused on small time-scales \citep[e.g., see][]{thomas2010,mcdermid2015}, and current CC data are found to be compatible with an exponentially delayed SFH with $\tau<0.3$ Gyr \citep{moresco2012,moresco2016}.
In Section~\ref{sec:Hzmeas} we provide an estimate of the total systematic covariance taking into account all components, including the uncertainty on the SFH.

Synthetic spectra have been extracted using the \verb|Galaxev| suite of codes for the BC16, BC03, and M11 models\footnote{Available at \url{http://www.bruzual.org/}.}; the online web tool\footnote{Available at \url{http://research.iac.es/proyecto/miles/pages/webtools/tune-ssp-models.php}.} for E-MILES; and the Python version of \verb|FSPS| for the FSPS models\footnote{Available at \url{http://dfm.io/python-fsps/current/}.}.
The complete library of the synthetic models created is made publicly  available\footnote{Available at \url{https://gitlab.com/mmoresco/library_syntheticspectra_CC}.}.

\subsection{Construction of the $D4000_n-$Age Relations}
\label{sec:measD4000}

\begin{figure*}[t!]
	\centering
	\includegraphics[width=0.85\textwidth]{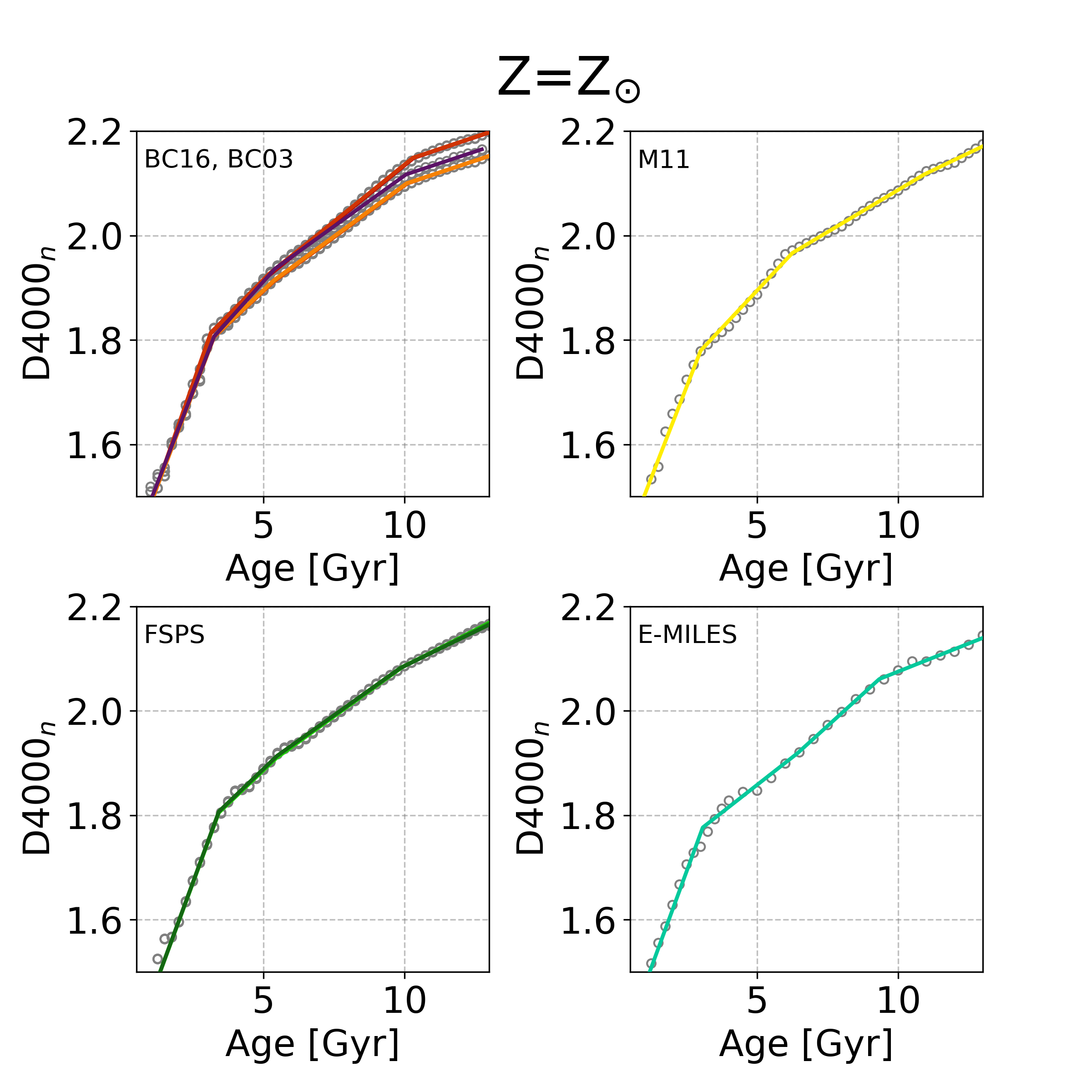}
	\caption{Solar metallicity $D4000_n-$age relation for the SSP models considered in our library, described in Table~\ref{tab:tab1}. The gray points represent values measured on the models, while the colored lines the piecewise linear fit to the data. The four panels show the relation for the BC16/BC03 models (upper left panel), M11 model (upper right panel), FSPS model (lower left panel), and E-MILES (lower right panel).}
	\label{fig:Dnage2}
\end{figure*}

The $D4000$ is a spectral feature that appears in galaxy spectra at 4000~\AA~restframe as a break generated by the contribution of several absorption features (the most prominent being the \ion{Ca}{2} H and K lines). This feature is defined as the ratio of the flux $F_{\nu}$ below and above the break, and, depending on the width of the windows considered, it is possible to define a wide $D4000$ \citep[$D4000_w$, 3750~\AA~$<\lambda<$ 3950~\AA~and 4050~\AA~$<\lambda<$ 4250~\AA, ][]{bruzual1983} and a narrow $D4000$ \citep[$D4000_n$, 3850~\AA~$<\lambda<$ 3950~\AA~and 4000~\AA~$<\lambda<$ 4100~\AA, ][]{hamilton1985}. Compared to other absorption features, the $D4000$ has the advantage of being easily measured; it does not require a particularly high spectral resolution to be detected.
For consistency with previous analyses obtained with the CC method, here we adopt the narrow definition $D4000_n$, since it has been demonstrated to be less dependent on reddening effects \citep{balogh1999}.

We measure the $D4000_n$ on all synthetic spectra of the library discussed in Section \ref{sec:lib}, constructing the $D4000_n-$age relations. Figure \ref{fig:Dnage2} shows these relations for the case of interest of solar metallicity. Different lines correspond to different models as described in Table~\ref{tab:tab1}. 
In fact, there is an extensive literature finding that massive and passively evolving galaxies, from the local universe up to $z\approx2$, have solar to slightly oversolar metallicities \citep{gallazzi2005,onodera2012,gallazzi2014, conroy2014,onodera2015,mcdermid2015,citro2016,comparat2017,estrada2019,kriek2019,morishita2019,saracco2019}. In the following, we will, therefore, focus our analysis only on solar metallicity models, since they are, among the available ones, the most representative for this population. Nonetheless, the effects of metallicity are then discussed in detail in Sec.~\ref{sec:met}.

As a final step, we need to use the previously measured $D4000_n-$age relations to simulate a ``mock'' measurement of $\Delta z/\Delta D4000_n$.
In order to do it, given a redshift interval $\Delta z$ identified by a pair of redshifts $z_1$ and $z_2$, we have to estimate the age of the corresponding galaxy population that we are aiming to simulate at these redshifts. From a fiducial cosmology, we can easily relate the age of a galaxy to its redshift with the relation
\begin{equation}
{\rm age}(z)={\rm age}_U(z)-{\rm age}(z_f)
\label{eq:agez}
\end{equation}
where ${\rm age}_U(z)$ is the age of the universe at the given redshift, ${\rm age}(z_f)$ is the age at which the galaxy population is formed, and $z_f$ is its corresponding formation redshift\footnote{We estimate the quantity $age(z)$ with the publicly available Python libraries for cosmological calculations \texttt{CosmoBolognaLib} \citep{marulli2016}}.
Given this equation, we are able to connect the redshift to the age of a simulated galaxy population and, with the $D4000_n-$age relation obtained (for a given model), the age of the measured $D4000_n$. In this way, we can associate with the pair of redshifts previously discussed a pair of $D4000_{n,1}$ and $D4000_{n,2}$, from which we derive $\Delta D4000_{n}$; this constitutes our ``mock''  $\Delta z/\Delta D4000_n$ from Section~\ref{sec:method}.
In Section~\ref{sec:measmatrix} we assess the impact of considering different ranges of formation redshifts in the analysis compatible with the properties of our chronometers. We anticipate here, however, that the assumption of a formation redshift $z_f$ is needed in the analysis just to relate the redshift of a population to a simulated $D4000_n$, and we verified that the results do not depend significantly on the specific choice.

\subsection{Measurement of the Slope of the $D4000_n-$Age Relations}
\label{sec:measslope}

To obtain the slope of the $D4000_n-$age relations of Section~\ref{sec:measD4000}, i.e. the parameter $A$ in Equation~\ref{eq:Hz_D4000}, different approaches can be exploited, from directly measuring the local slope to smoothing the curves to minimize the impact of small fluctuations in the relations (see Figure~\ref{fig:Dnage2}). 

Here we explore two different methods. We measure the slope assuming a linear relation over different ranges of $D4000_n$; we will refer to this as the {\it piecewise linear slope}. Alternatively, we fit the slope as a function of $D4000_n$, with a polynomial relation or a cubic spline; we will refer to this as the {\it interpolated slope}.
Since we are dealing with the estimate of a derivative, these two methods allow us to minimize the impact of small variations in the $D4000-$age relation that can have a large and random impact on the estimate of $H(z)$.
For completeness, we have also performed the analysis estimating the {\it local slope} and propagated this measurement following the same workflow as in Figure~\ref{fig:workflow}. We find the piecewise linear slope approach to be the most robust; hence, we only discuss the other approaches in Appendix \ref{sec:intpslope}.

\paragraph{Piecewise linear slope.} 

\begin{table*}[t!]
	\begin{center}
		\caption{Parameters of the Piecewise Linear Slope of the $D4000_n-$Age Relations for Solar Metallicities as Discussed in Section~\ref{sec:measslope}}
		\begin{tabular}{cccccccc}
			\hline
			\hline
			&  &  &  & Slope & Slope & Slope & Slope\\
			& $D4000_{n}$ & $D4000_{n}$ & $D4000_{n}$ & Lower & Medium-lower & Medium-higher & Higher\\
			Model & First Break & Second Break & Third Break & $D4000_{n}$ Range & $D4000_{n}$ Range & $D4000_{n}$ Range & $D4000_{n}$ Range \\
			\hline
			BC16, miles, chab & 1.813 & 1.922 & 2.148 & 0.154 & 0.054 & 0.044 & 0.0186\\
			BC16, miles, kroup & 1.814 & 1.926 & 2.148 & 0.154 & 0.054 & 0.044 & 0.0180\\
			BC16, miles, salp & 1.814 & 1.926 & 2.149 & 0.154 & 0.055 & 0.044 & 0.0181\\
			BC16, stelib, chab & 1.805 & 1.911 & 2.099 & 0.142 & 0.050 & 0.040 & 0.0183\\
			BC16, stelib, kroup & 1.807 & 1.914 & 2.010 & 0.142 & 0.051 & 0.039 & 0.0178\\
			BC16, stelib, salp & 1.806 & 1.914 & 2.010 & 0.142 & 0.051 & 0.040 & 0.0179\\
			M11, miles, chab & 1.780 & 1.967 & 2.117 & 0.139 & 0.058 & 0.032 & 0.0260\\
			FSP,  miles, chab & 1.805 & 1.909 & 2.083 & 0.147 & 0.051 & 0.039 & 0.0277\\
			FSPS, miles, kroup & 1.807 & 1.913 & 2.082 & 0.147 & 0.052 & 0.039 & 0.0269\\
			FSPS, miles, salp & 1.807 & 1.914 & 2.081 & 0.147 & 0.052 & 0.038 & 0.0265\\
			Vazd, emiles, chab & 1.763 & 1.807 & 2.074 & 0.145 & 0.058 & 0.044 & 0.0202\\
			BC03, stelib, chab & 1.807 & 1.936 & 2.117 & 0.140 & 0.060 & 0.039 & 0.0179\\
			\hline
		\end{tabular}
		\label{tab:slopes}
	\end{center}
\end{table*}

This approach is used to estimate $H(z)$ with the CC method in several works \citep{moresco2012,moresco2015,moresco2016}. It is based on the fact that, at fixed stellar metallicity, the $D4000_n-$age relations are extremely well reproduced \citep[as demonstrated in][]{moresco2012} by a simple linear fit, once the $D4000_n-$age curve is divided into appropriate $D4000_n$ ranges that take into account the knees of the relations. The origin of the linear piecewise behavior of $D4000$ as a function of age can be traced back to the behavior of $D4000$ as a function of the spectral type of single stars or, equivalently, of their effective temperature. \citet{bruzual1983} already found a clear dependence on $D4000$ as a function of spectral type, where at least three slopes can be clearly identified: one for stars from O to A type, one for stars from A to G type, and one for stars for types G to K-M (see his Figure 3). A similar result is obtained by \citet{gorgas1999}, where the analysis of the $D4000$ as a function of  $\theta=5040/T_{\rm eff}$ clearly highlights also in this case at least three regimes, $\theta\lesssim0.6$, $0.6\lesssim\theta\lesssim0.8$, and $\theta\gtrsim0.8$, corresponding to effective temperature $T_{\rm eff}\approx$ 8400 and 6300 K (see their Figures 4 and 5). The aging of a stellar population, as also highlighted in \cite{bruzual1983}, can be mapped in an evolution of the stars composing it, shifting from a population dominated by younger and hotter to older and colder stars; this will result in the piecewise $D4000$ relation that we observe in Figure~\ref{fig:Dnage2}, while the exact mixing of spectral types will depend on the ingredients used in the model.

The advantage of estimating the slope with this method is that it is not strictly tied to the exact measured value of $D4000_n$, as are the local or interpolated slopes. On the contrary, given only the range to which a particular $D4000_n$ belongs, it provides a unique value of the slope $A$. Being independent of a particular measured value of $D4000_n$, it therefore maximizes the strength of the differential form of Equation~\ref{eq:Hz_D4000}, where the constraints on $H(z)$ are obtained just from the measurement of a differential age (or $D4000_n$) evolution of CCs.

To perform the piecewise linear fit to the $D4000_n-$age relations, we adopt the public Python code \verb|pwlf|\footnote{\url{https://pypi.org/project/pwlf/}.} \citep{pwlf}.
Here we assume that we have three different breaks in the $D4000_n-$age relations in each model, as done in previous analyses \citep{moresco2012, moresco2016}, although, as can be seen in Figure~\ref{fig:Dnage2}, for some models, the presence of three breaks is more evident by eye than for others. The motivation of this assumption is the fact that we consider each of these models to be a particular representation of the same underlying truth, and therefore we decide to adopt the same number of breaks for each fit. In this way, we divide the $D4000_n-$age relations into four ranges, to which we will refer in the following as lower, medium-lower, medium-higher, and higher $D4000_n$ ranges.
The results are shown in Table~\ref{tab:slopes}, where we report both the values of the slopes in the various ranges and the position of the breaks.

\begin{figure*}[t!]
	\centering
	\includegraphics[width=1.\textwidth]{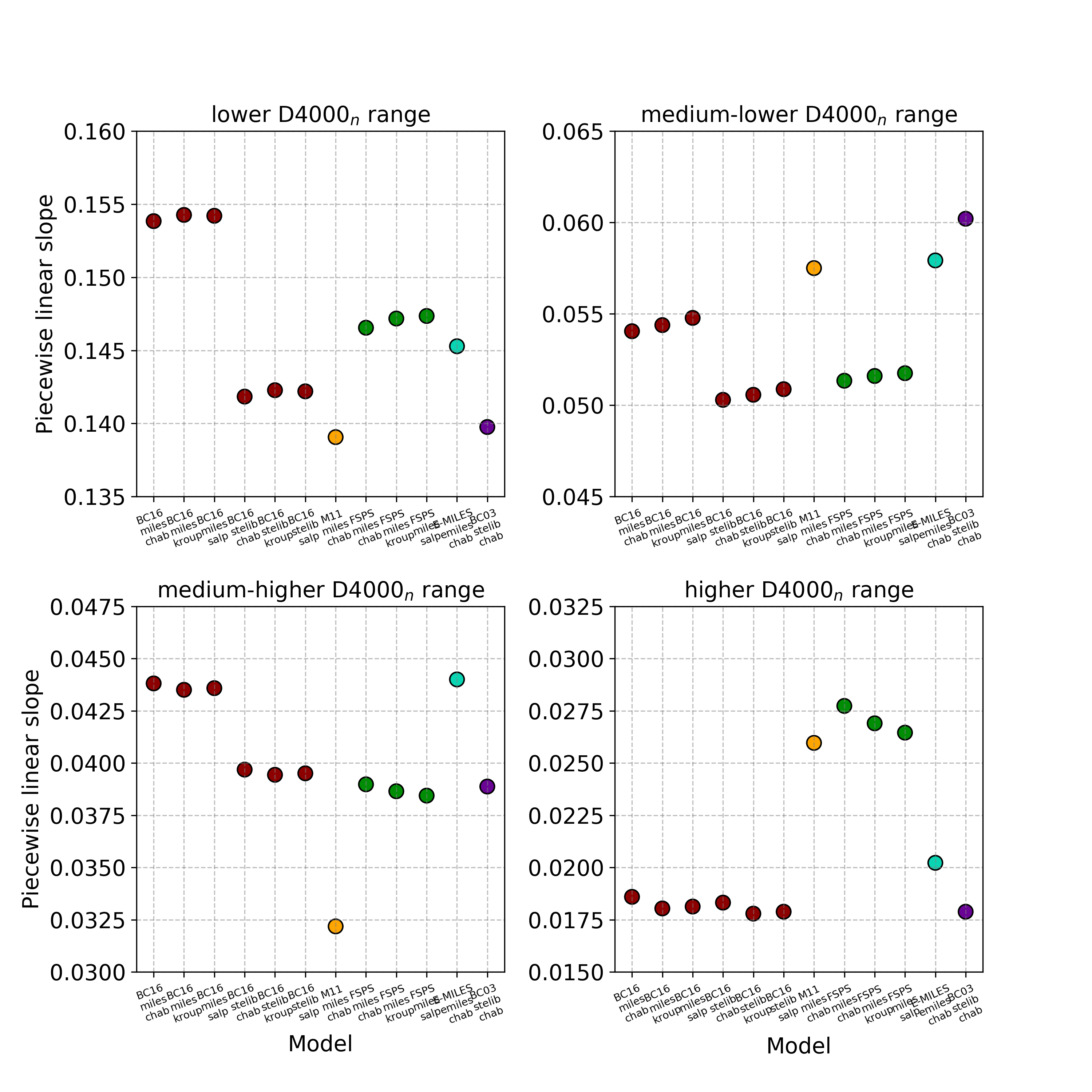}
	\caption{Piecewise linear slopes of the $D4000_n-$age relations for solar metallicities estimated for all of the various models in the different $D4000_n$ regimes discussed in Section~\ref{sec:measslope}. Measurements obtained with the same SPS model have been represented with the same color to ease the comparison.}
	\label{fig:slopes}
\end{figure*}

We find the positions of the breaks (a free parameter in the fit) to be very consistent among different models, with a first break around $D4000_n=1.8$, a second one around $D4000_n=1.9$, and a third one around $D4000_n=2.1$, in agreement with the values adopted in previous analyses. Moreover, we have verified the goodness of the piecewise linear fit for all models by estimating the coefficient of determination\footnote{For a linear fit the coefficient of determination is the square of the Pearson correlation coefficient.} $r^2$.
Checking all models, we find an average value of $\langle r^2\rangle =0.987\pm 0.014$ (with a minimum value of 0.941 and a maximum value of 0.999), indicating that a linear fit is well motivated.

Figure~\ref{fig:Dnage2} shows, for each of the $D4000_n$ regimes, the raw $D4000_n-$age relation as directly measured from the models in Section~\ref{sec:measD4000} (gray points) and the linear fits for the different models (colored lines) at solar metallicity in the various ranges. 
The slope values (for solar metallicity) are visualized in Figure~\ref{fig:slopes}, where models drawn from the same SPS are shown with the same color. It is interesting to note that the impact of different IMFs on the slopes $A$ is subdominant compared to the other ingredients considered. 
On the other hand, the effects of the SPS model and the stellar libraries are the most important. How this propagates to the $H(z)$ measurement is quantified in Section~\ref{sec:measmatrix}.

\subsection{Bias Due to the Choice of SPS Model}
\label{sec:measmatrix}

Finally, to estimate the impact of adopting different SPS models on $H(z)$, we combine all previously described steps as shown in the workflow of Figure~\ref{fig:workflow}.
Here we assume that stellar metallicity is determined with negligible error; the systematic effect due to an error in the measurement of the metallicity is assessed later in Section~\ref{sec:met}.

\begin{figure*}
\centering
\includegraphics[width=0.48\textwidth]{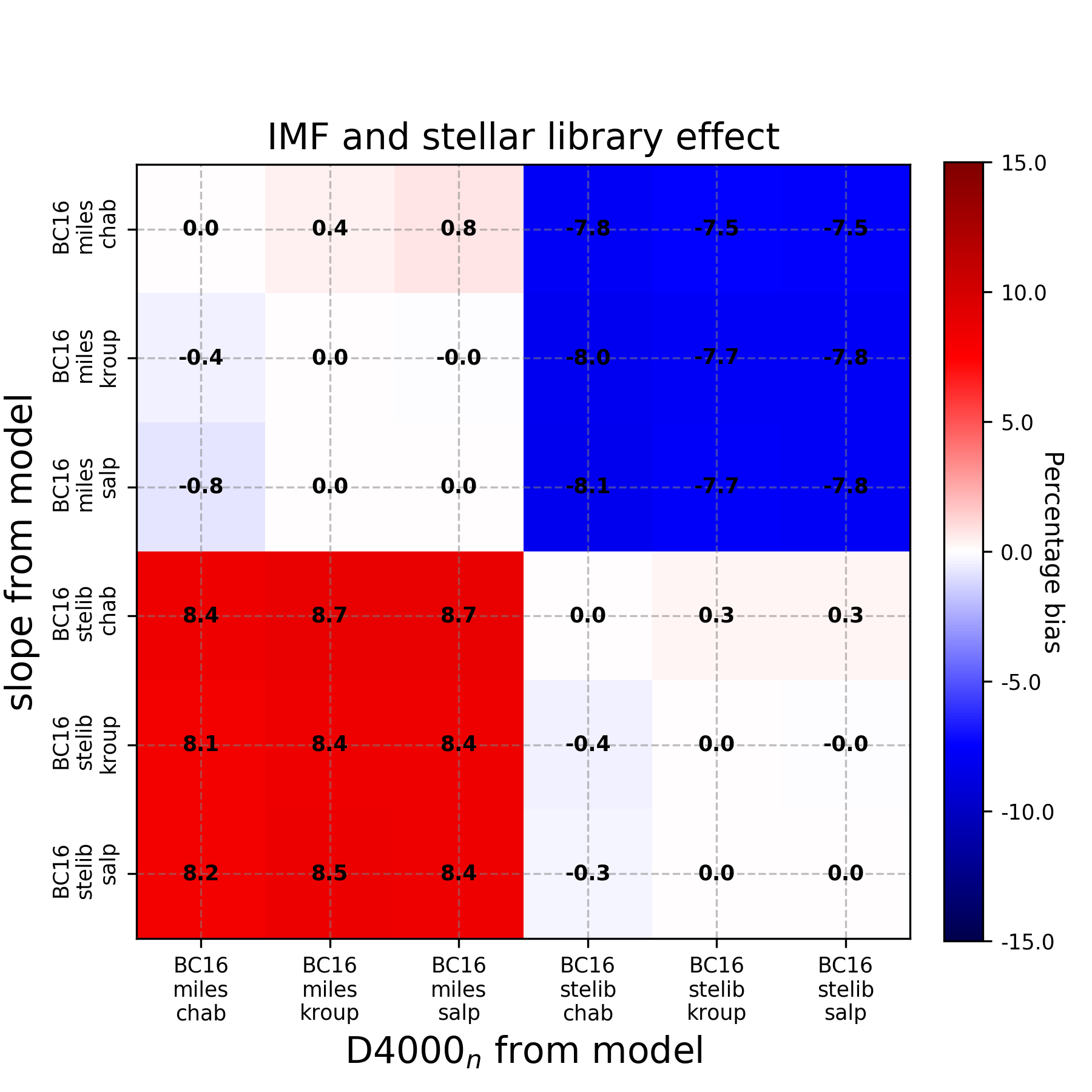}
\includegraphics[width=0.48\textwidth]{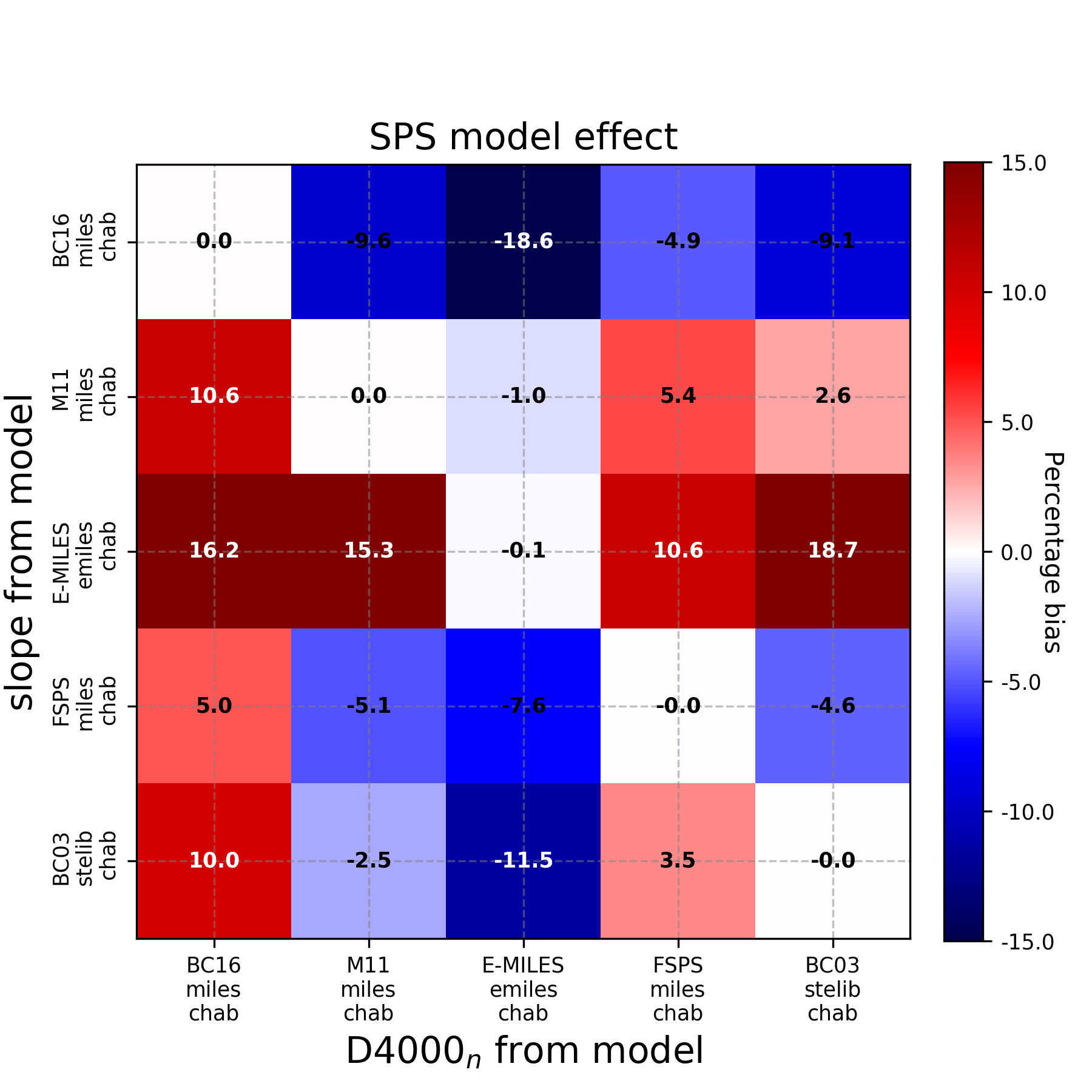}
\caption{Matrix of the $H(z)$ percentage difference as defined in Equation~\ref{eq:etap}, where the x-axis label indicates from which model the $D4000_n$ has been extracted, and the y-axis indicates from which model the slope $A$ (see Equation~\ref{eq:Hz_D4000}) has been extracted, considering a piecewise linear slope as discussed in the text. 
For illustrative purposes, we only show the matrix for  $1.00<z<1.05$. In the left panel, we compare all models extracted from the BC16 SPS models, with different IMFs and stellar libraries as indicated in the caption, and in the right panels are shown models obtained with the same Chabrier IMF but from different codes.  We also show the outdated BC03 models as a reference and not because they are being used in the current analysis, but they are an example of the level of model convergence and for reference for previous works.}
\label{fig:eta}
\end{figure*}

The general procedure is outlined below.
\begin{enumerate}
\item We consider redshifts pairs $(z_1,z_2)$ from which we derive $\Delta z=z_2-z_1$; the effective redshift $z$ is defined as the mean of $z_1$ and $z_2$. Unless otherwise stated, we set $\Delta z=0.05$ and sample redshifts up to $z_{max} =1.5$.
\item Assuming our fiducial cosmology and a formation redshift $z_f$ (drawn from the range $1.5<z_f<5$, as justified in the following), the previous redshifts are converted in ages $\rm (age_1,age_2)$ following Equation~\ref{eq:agez}.
\item For each model (labeled by running index $a$), the ages are converted to values ($D4000_{n,1},D4000_{n,2}$) from which we derive $\Delta D4000_{n\,a}=D4000_{n,2}-D4000_{n,1}$, where the $D4000_n$ values are drawn from the linearized $D4000_n-$age relations shown in Figure~\ref{fig:Dnage2}, to smooth the small oscillations in the relations discussed in Section~\ref{sec:measD4000}. We limit our analysis to the range of $D4000_n$ values probed by observations, i.e. $1.5<D4000_n<2.1$ \citep{moresco2012,moresco2015,moresco2016}, and, therefore, given a redshift and a formation redshift as described in step (2), our simulated measurements are constructed to satisfy this condition.
\item The slope $A_b$ is then obtained for all the models (labeled by running index $b$) according to the values of $D4000_n$, considering a piecewise linear slope. 
If $D4000_{n,1}$ and $D4000_{n,2}$ happen to be across a knee, it is not straightforward to assign a slope. In Appendix~\ref{sec:slope_btw_knees} we also provide the formalism we adopt to estimate the slope in this condition.
\item Combining steps (3) and (4), we obtain $H(z)$ as
\begin{equation}
H(z)_{a,b}=-\frac{A_b}{1+z}\frac{\Delta z}{\Delta D4000_a}\,.\label{eq:Hz_ab}
\end{equation}
\end{enumerate}
Note that the first index $a$ in $H(z)_{a,b}$ corresponds to the model used to generate the ``mock'' $D4000_n$ measurements and the corresponding $\Delta z/\Delta D4000_n$, while the second index $b$ refers to the model used to provide the slope $A$ of the $D4000_n-$age relation. 

From Equation~\ref{eq:Hz_ab}, it is possible to define the relative  bias due to SPS modeling as
\begin{equation}
\eta(z)_{a,b}=\frac{H(z)_{a,b}-H_{\rm fid}(z)}{H_{\rm fid}(z)}
\label{eq:etap}
\end{equation}
where $H_{\rm fid}(z)$ is the Hubble parameter for the assumed fiducial cosmology at a given redshift.
In the following, it is also useful to define the percentage bias {\boldmath$\eta$} = $\eta\times100$.
Equation \ref{eq:etap} allows us to quantify the error induced on the estimate of $H(z)$ by taking the parameter $A$ and the quantity $dD4000_n$ from two different (linearized) models. 

We have verified that, with the correction introduced in Appendix~\ref{sec:slope_btw_knees}, $\eta(z)_{a,a}=0$ to subpercent level, and therefore that generating a $D4000_n$ measurement and fitting it with the same model exactly reproduces the expected fiducial $H(z)$.
It is also important to notice that the matrix $\eta$ is not symmetrical, since a $\Delta D4000_n$ extracted from model $a$ and fitted with model $b$ would not give the same estimated $H(z)$ when extracting the $\Delta D4000_n$ from model $b$ and fitting it with model $a$; therefore, $\eta_{a,b}\neq\eta_{b,a}$. 

In particular, (i) to assess the impact of the IMF, we consider the available models at fixed SPS and stellar library, namely BC16 with MILES, BC16 with STELIB, and FSPS with MILES (varying the IMF between Chabrier, Kroupa, and Salpeter); (ii) to quantify the impact of the stellar library, we consider the available models at a fixed IMF and SPS, namely BC16 with Chabrier, BC16 with Kroupa, and BC16 with Salpeter (varying the stellar library between MILES and STELIB); and (iii) to estimate the impact of the adopted SPS, we consider the available models at a fixed IMF and stellar library, namely BC16, M11, FSPS, and E-MILES, all with MILES and a Chabrier IMF.

To assess a possible dependence on the formation redshift $z_f$, at each redshift and for each model combination, we consider a range $z_f=[1.5-5]$, in agreement with observational constraints \citep{franx2003,cimatti2004,mccharty2004,daddi2005,treu2005,renzini2006,pozzetti2010,thomas2010,mcdermid2015,carnall2018,carnall2019}. However, we also explore other less and more conservative choices, namely $z_f=[1.5-3]$ and $[1.5-10]$. As already discussed in Section~\ref{sec:method}, we find that our results do not depend significantly on the choice of the grid of $z_f$, as we will discuss in Section~\ref{sec:results}. 

Each element of the matrix $\eta(z)_{a,b}$ in a given redshift bin is obtained by estimating the median of the results for the range of formation redshifts considered. A typical result for the {\boldmath $\eta$} matrix at $1.00<z<1.05$ is shown in Figure~\ref{fig:eta}. The other redshift bins show a similar behavior, and the results of the analysis of all redshift bins are presented in the following section.

\subsection{Bias Due to the Uncertainty on Stellar Metallicity}
\label{sec:met}
The impact of the uncertainty on the estimate of stellar metallicity on $H(z)$ is estimated with a similar procedure to the one discussed in Section~\ref{sec:measmatrix}. In this case, we take advantage of the \verb|FSPS| code, which allows one to simulate
spectra directly with a user-defined stellar metallicity, interpolating between existing ones. In this case, we simulate uncertainties of $\pm$10\%, $\pm$5\%, and $\pm$1\% around the solar metallicity. This range allows us to probe both a conservative estimate of the expected error on metallicity \citep[a similar error bar is obtained, e.g., in][from a full spectral fitting of BOSS spectra with different independent codes and SPS models]{moresco2016} and an ideal case where future improvements in modelization and analysis would constrain metallicity at the 1\% level, to forecast how much this systematic uncertainty could be narrowed down.

\section{Results}
\label{sec:results}
Analazing the bias matrices, we can appreciate that they are approximately
antisymmetrical, $\eta_{a,b}\sim-\eta_{b,a}$.

We also notice that the behavior of $\eta_{a,b}$ with redshift is random; i.e., as a function of redshift, no model systematically overpredicts or underpredicts the Hubble parameter with respect to another. This is due to the fact that by changing the redshift range, we are changing also the $D4000_n$ values spanned by the data, and the slope of the $D4000_n-$age relations changes too,  not monotonically, but following the relations shown in Figure~\ref{fig:slopes}. 

To quantify the overall systematic error on $H(z)$, we consider all of the available model combinations separately for the IMF, stellar library, and SPS contribution (as shown in Figure~\ref{fig:eta}), estimating for each redshift bin the mean of the absolute value of the elements of the matrix $\eta_{a,b}$, computing the quantity

\begin{equation}
\widehat{\eta}(z)={\rm mean}({\rm abs}(\eta(z)_{a,b}))\,.
\label{eq:meaneta}
\end{equation}

We also estimate $\eta^{\rm max}$, the maximum of the absolute value of the elements per each redshift bin, as a maximum catastrophic error one would do if the real $D4000_n-$age relation follows a particular model, and it is fitted with the most discordant one. We find that among all redshift bins, it varies from a minimum of $\sim$10\% up to peaks of $\sim$35\%, being, on average, $\sim$25\% over all redshift bins. These values, however, represent a strict upper limit to the error.

When studying the dependence of the $\eta_{ab}$ on the SPS model (for a fixed IMF and stellar library), we note that at each redshift, there is one model that is discrepant from the other three; however, it is not always the same model at every redshift. This effect could be also better visualized in Figure~\ref{fig:slopes}, where it is evident how in each $D4000_n$ range, there are outliers.
For this reason, we have decided to also present the results of the estimated bias obtained by excluding the most discrepant model at each redshift, which we refer to as ``odd one out,'' with a procedure similar to a sigma-clipping. This choice is motivated by the rationale that all models should be theoretical representations of the same underlying truth; therefore, either the majority of the models are better calibrated on data and the discordant one should be revised, or the single discordant one is correct and all of the others require improvements. In either case, the associated systematic error estimated without considering the outlier could be considered as an improved estimate.

\begin{figure}[t!]
	\centering
	\includegraphics[width=0.5\textwidth]{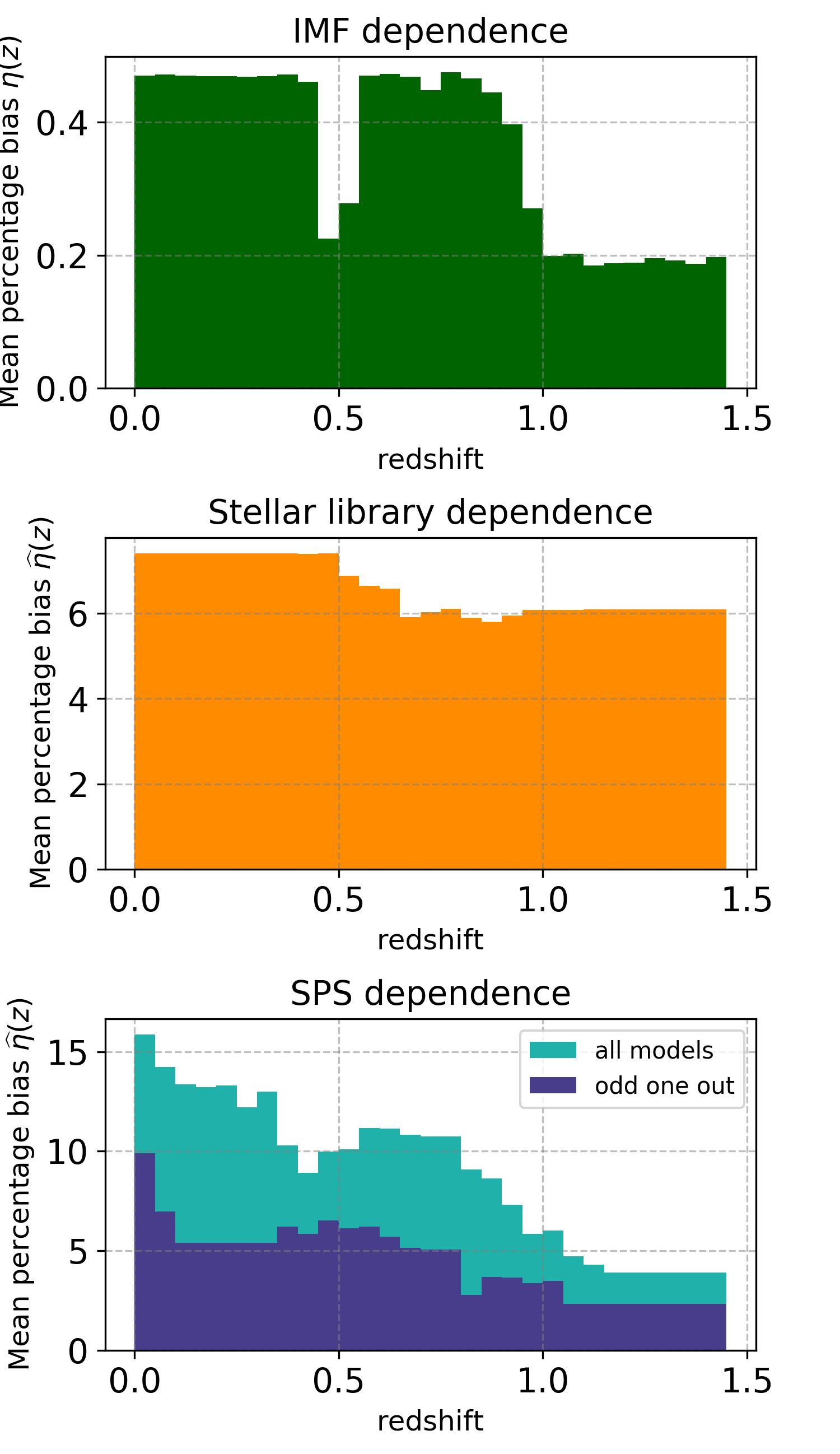}
	\caption{Mean percentage bias $\widehat{\eta}(z)$ on $H(z)$ (as defined in Equation~\ref{eq:meaneta}) as a function of redshift, averaged over the model combinations shown in Figure~\ref{fig:eta} for the piecewise linear slope case. In the bottom panel (darker color), we show the uncertainty when the odd model is left out (see text and Table~\ref{tab:res_mean} for more details).}
	\label{fig:offset}
\end{figure}

\begin{table}[t!]
	\begin{center}
		\caption{Mean Percentage Bias $\widehat{\eta}(z)$ as a Function of Redshift.}
		\begin{tabular}{ccccc}
			\hline
			\hline
			& \% Offset & \% Offset & \% Offset & \% Offset \\
			$z$ & IMF & St. Lib. & SPS Model & SPS Model\\
			& & & & (Odd One Out) \\
			\hline
			0.075 & $0.47$ & $7.40$ & $15.86$ & $9.91$ \\
			0.125 & $0.47$ & $7.40$ & $14.23$ & $6.98$ \\
			0.175 & $0.47$ & $7.40$ & $13.34$ & $5.40$ \\
			0.225 & $0.47$ & $7.40$ & $13.21$ & $5.40$ \\
			0.275 & $0.47$ & $7.40$ & $13.29$ & $5.40$ \\
			0.325 & $0.47$ & $7.40$ & $12.20$ & $5.40$ \\
			0.375 & $0.47$ & $7.40$ & $12.99$ & $5.40$ \\
			0.425 & $0.47$ & $7.40$ & $10.29$ & $6.20$ \\
			0.475 & $0.46$ & $7.39$ & $8.91$ & $5.86$ \\
			0.525 & $0.23$ & $7.40$ & $9.99$ & $6.51$ \\
			0.575 & $0.28$ & $6.87$ & $10.09$ & $6.12$ \\
			0.625 & $0.47$ & $6.65$ & $11.17$ & $6.21$ \\
			0.675 & $0.47$ & $6.57$ & $11.12$ & $5.71$ \\
			0.725 & $0.47$ & $5.90$ & $10.81$ & $5.16$ \\
			0.775 & $0.45$ & $6.03$ & $10.75$ & $5.05$ \\
			0.825 & $0.47$ & $6.10$ & $10.75$ & $5.05$ \\
			0.875 & $0.47$ & $5.89$ & $9.08$ & $2.79$ \\
			0.925 & $0.44$ & $5.80$ & $8.62$ & $3.70$ \\
			0.975 & $0.40$ & $5.94$ & $7.32$ & $3.65$ \\
			1.025 & $0.27$ & $6.07$ & $5.84$ & $3.37$ \\
			1.075 & $0.20$ & $6.08$ & $6.02$ & $3.49$ \\
			1.125 & $0.20$ & $6.07$ & $4.72$ & $2.33$ \\
			1.175 & $0.19$ & $6.09$ & $4.31$ & $2.33$ \\
			1.225 & $0.19$ & $6.09$ & $3.90$ & $2.33$ \\
			1.275 & $0.19$ & $6.09$ & $3.90$ & $2.33$ \\
			1.325 & $0.20$ & $6.09$ & $3.91$ & $2.34$ \\
			1.375 & $0.19$ & $6.09$ & $3.90$ & $2.34$ \\
			1.425 & $0.19$ & $6.09$ & $3.90$ & $2.33$ \\
			1.475 & $0.20$ & $6.09$ & $3.91$ & $2.34$ \\
			\hline
		\end{tabular}
		\label{tab:res_mean}
	\end{center}
\end{table}

The results are reported in Table~\ref{tab:res_mean} and Figure~\ref{fig:offset}. In Table~\ref{tab:res_tot} we have also averaged the results as a function of redshift to provide an average percentage error. 
We find the following results.
\begin{itemize}
\item Dependence on IMF. The effect of an IMF variation has the smallest impact on the $H(z)$ measurement, with a mean percentage error $<$0.5\% as a function of redshift and an average value of 0.4\%.
\item Dependence on stellar library. The effect of the different stellar libraries considered is, instead, more relevant, with a mean value $\sim$6.5\%, almost independent of the redshift.
\item Dependence on SPS model. The effect of considering different SPS models impacts the $H(z)$ measurement with a mean percentage bias of $\sim$9\%, on average, with a decreasing trend with increasing redshift between $z=0$ and $1.5$. We can find the explanation of this behavior by looking at Figure~\ref{fig:slopes}, which shows the slopes of the $D4000_n-$age relations for the different models.
The percentage difference between the slopes of different models is higher at higher values of $D4000_n$ (and hence at higher ages and lower redshifts) and smaller at smaller values of $D4000_n$ (that dominate the higher redshifts). Moreover, it is important to stress that the real power of the CC method is at $z>0.2$, where there is enough volume to observe enough passively evolving galaxies. This is the case to date and will be even more so with future data. In this redshift range, $\widehat{\eta}(z)\lesssim13$\%.
\item Dependence on SPS model adopting an odd-one-out approach. As discussed above, the error due to different SPS models adopted is in many cases mostly driven by a single model significantly different from the others. By excluding the odd model out, we find that the errors are significantly reduced to an average value $\sim$4.5\% and smaller than 6\% for $z>0.2$. This is interesting, since it shows a path to further reduce systematic uncertainties through a more concerted and comprehensive effort of model comparison to highlight the strengths and weaknesses of each model and possibly lead to more convergent models.
\item Dependence on the formation redshift $z_f$. Finally, we analyze the impact of changing the grid of $z_f$ in the analysis, exploring the possibilities discussed in Section~\ref{sec:measmatrix} and find that it does not significantly affect the results. In particular, we find differences smaller than 0.5\%, on average, for the percentage offsets due to the IMF and stellar library and smaller than 3\%, on average, for the percentage offsets due to SPS models (smaller than 1.5\% for the SPS models accounting for the odd-one-out option).
\end{itemize}

\begin{table}
\begin{center}
\caption{Average Systematic Impact on $H(z)$ Determination Due to SPS Modeling (Upper Part of the Table) and to Uncertainty on Stellar Metallicity (Lower Part of the Table) on the CC Approach.}
\begin{tabular}{rccc}
\hline
\hline
\multicolumn{4}{c}{Total Systematic Error Budget on $H(z)$ with the CC Method}\\
\multicolumn{1}{r}{} & \multicolumn{1}{c}{Min} & \multicolumn{1}{c}{Max} & \multicolumn{1}{c}{Mean}\\
\hline
{Effect due to models} &  &  & \\
IMF & 0.19\% & 0.47\% & 0.36\% \\
Stellar library & 5.80\% & 7.40\% & 6.57\%\\
SPS model  & 3.90\% & 15.86\% & 8.91\% \\
SPS model (odd one out)  & 2.33\% & 9.91\% & 4.53\% \\
\hline
{Effect due to metallicity} &  &  & \\
(10\% error) & 6.01\% & 19.33\% & 9.02\% \\
(5\% error) & 1.87\% & 6.97\% & 4.16\% \\
(1\% error) & 0.18\% & 0.77\% & 0.49\%  \\
\hline
\end{tabular}
\end{center}
{\textbf{Note.} We report the minimum, the maximum, and the mean bias as a function of redshift, separately for each component (IMF, stellar library, SPS model).}
\label{tab:res_tot}
\end{table}

\begin{figure*}[t!]
\centering
\includegraphics[width=0.99\textwidth]{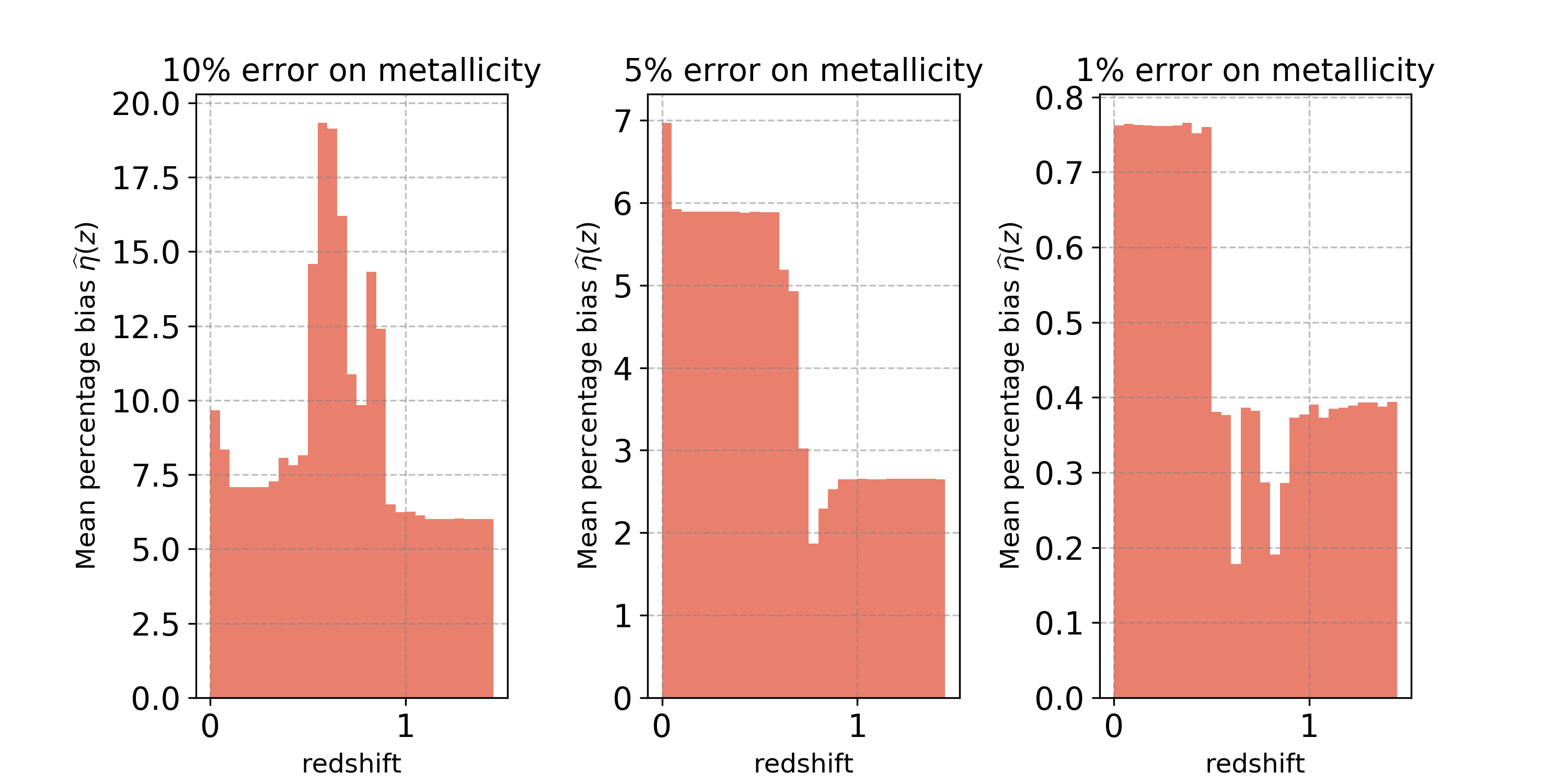}
\caption{Impact of metallicity uncertainty on the percentage bias as a function of redshift. The different figures show how the percentage bias changes (assuming a piecewise linear slope) for a given error in the measured metallicity of $\pm$10\%, $\pm$5\%, and $\pm$1\%.}
\label{fig:offset_met}
\end{figure*}

\begin{figure}[t!]
	\centering
	\includegraphics[width=0.45\textwidth]{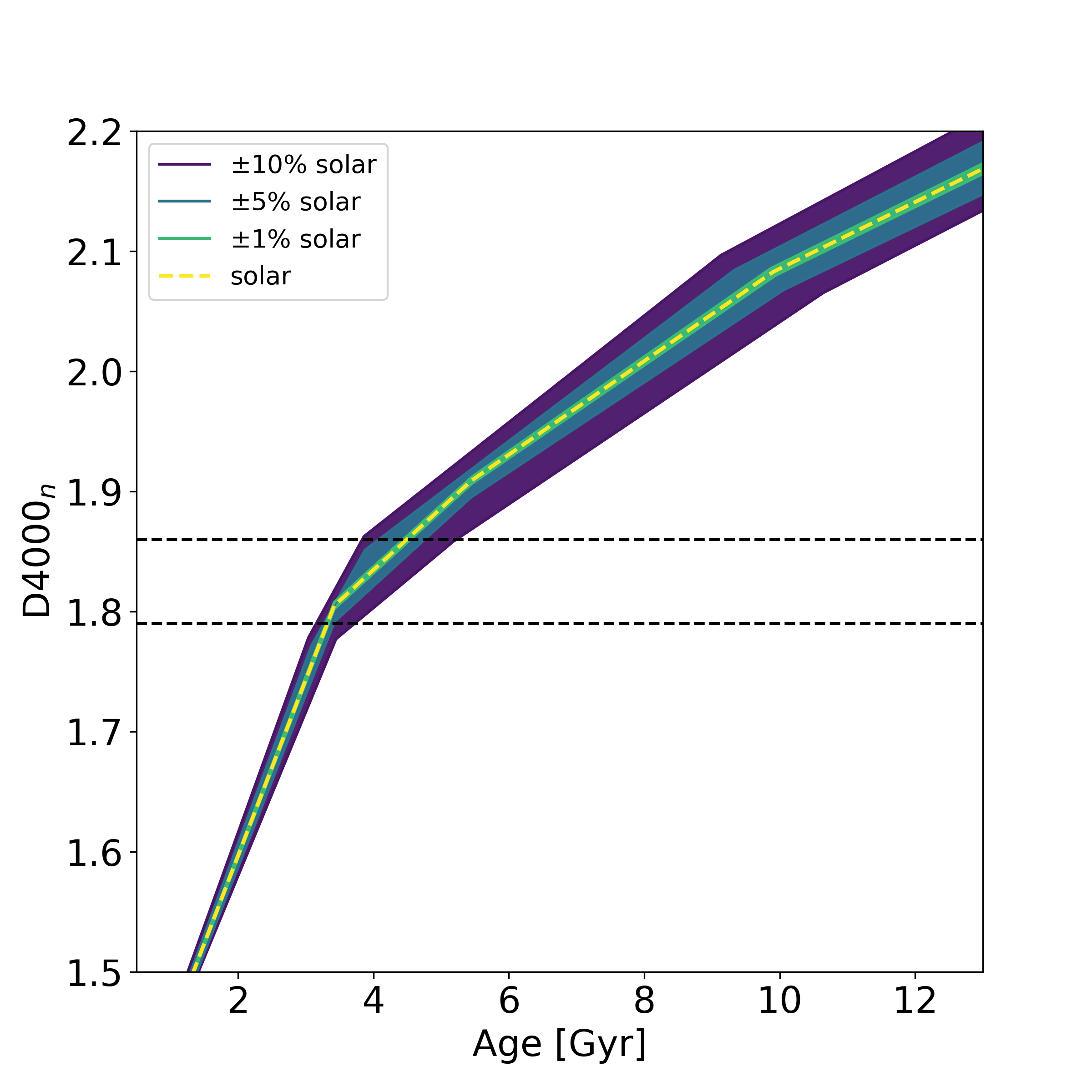}
	\caption{The $D4000_n-$age relation obtained with the FSPS models at different metallicities. The yellow line shows the solar metallicity, while in violet, azure, and green are shown a $\pm$10\%, $\pm$5\%, and $\pm$1\% variations, respectively. The dashed lines show the range $1.79<D4000_{n}<1.86$, corresponding to the redshift range associated with a larger bias in the $\pm$10\% case.}
	\label{fig:D4n_met}
\end{figure}

As a cross-check, we have also estimated the same total bias using the local slope (which, as discussed in Section~\ref{sec:measslope}, has not been explicitly presented here). The instability in its estimate, due to an intrinsically noisier measurement of the slope on unsmoothed data, results, as expected, in larger estimated errors. 

We remark that we averaged the results over all possible combinations of models without making any assumption about the correctness of one model over another. Such considerations are beyond the scope of this paper.
The procedure presented here could be easily repeated or reinterpreted in light of new considerations on the accuracy of various SPS models. In this paper, we have explored the main models proposed and used in the literature currently and in the recent past. Looking forward, if, in the future, data will allow constraining the range of available models (or IMFs, or stellar libraries) even more, this analysis could be simply repeated, possibly lowering the variance and hence the uncertainty due to SPS modeling.

The current error budget has been estimated  assuming solar and close-to-solar abundances and does not apply to significantly different metallicities (e.g., half or twice solar abundances). In the next paragraph, we assess the additional term in the error budget due to an uncertainty on the stellar metallicity.

\paragraph{Dependence on Stellar Metallicity}

Analogously, as presented in Section~\ref{sec:met}, we study the possible bias due to stellar metallicity errors, and the results are presented in Figure~\ref{fig:offset_met} and in Table~\ref{tab:res_tot}.
Analyzing the figure, we find a different behavior of the percentage bias in the case of a $\pm$10\% metallicity variation compared to the $\pm$5\% and $\pm$1\% cases. The origin of this difference can be tracked down to its nontrivial dependence on a number of factors. As previously discussed, many elements can have an impact on it, in particular, the position of the breaks in the $D4000_{n}-$age relations, the slope of the $D4000_{n}-$age relations, and the fact that different formation redshifts map in different $D4000_{n}$ values.
This can be better understood by looking at the obtained $D4000_{n}-$age relations for different stellar metallicities shown in Figure~\ref{fig:D4n_met}.
From the figure, it is possible to see that for small metallicity variations, the resulting relations are very close to the solar one, with a quasi-linear regime of variations and a small change in the slope and in the position of the breaks. For larger metallicity variations, instead, the changes are larger, resulting in a different behavior of the impact on $H(z)$ shown in Figure~\ref{fig:offset_met}. In particular, we note that at redshifts in the range $0.55<z<0.75$, corresponding to a larger bias in the $\pm$10\% case, we are mapping $D4000_{n}$ values in the range $1.79<D4000_{n}<1.86$, corresponding to the range where the relations are more different.

We find that an uncertainty on the metallicity of $\sim$10\% affects the $H(z)$ with a percentage error of $\sim$9\%, and this error is lowered to $\sim4$\% ($\sim$0.5\%) when the uncertainty on the metallicity is of the order of 5\% (1\%). This result is consistent with the findings of \citet{moresco2016}, where the uncertainty due to metallicity was dominating the error budget (given to the large available statistics), and an uncertainty on the metallicity of $\sim$10\% propagated to an error on $H(z)$ of $\sim$10\%.

\subsection{The Full Covariance Matrix for CCs}
\label{sec:Hzmeas}
The quantities shown in Table~\ref{tab:res_mean} can be finally used to construct the covariance matrix due to model ${\rm Cov}_{ij}^{\rm model}$. We define
\begin{equation}
\label{eq:Covcombine}
{\rm Cov}_{i,j}^{X}=\widehat{\eta^X}(z_i)\times H(z_i) \times \widehat{\eta^X}(z_j)\times H(z_j)
\end{equation}
where $X$ stands for the IMF, stellar library, and SPS contributions, as in Equation~\ref{eq:Covmodel}. The various contributions due to covariance can be then be added together to build the total covariance matrix. Users who want instead to sum the systematic contributions linearly rather than quadratically could instead use 
$\eta^{\rm model}=\sum_X\eta^X$ and then use Equation~\ref{eq:Covcombine}.

We note here that in the context of an application of the CC method to real data, how to combine the errors provided in Table~\ref{tab:res_mean} in Equation~\ref{eq:Covcombine} will depend on the quality of the data. Poorer data quality in terms of spectral resolution and signal-to-noise ratio (S/N) will result in a more uncertain metallicity and SFH estimate and hence larger systematic errors. On the other hand, an extended model comparison based on high-quality and high-resolution data would make possible a better convergence between the models in the future and significantly reduce its impact on the systematic error.

In the next section, we provide a few examples of how these contributions combine and propagate to the systematic error on a measurement of $H(z)$ in three illustrative cases: best-case and worst-case scenarios and an application to current measurements.

\subsection{Estimating the Systematic Errors for CCs: A Worked Example} 
\label{sec:examples}
\paragraph{Best-case scenario} 

Let us assume a high-S/N and high-resolution measurement of the spectra of CCs at $z\sim0.8$. This can be obtained by exploiting the new high-resolution instruments, e.g., X-Shooter or VIMOS/HR-Red. As an example, in the LEGA-C survey \citep{straatman2018}, they obtained spectra of single passive galaxies with a resolution $R\sim2500$ and a mean S/N=20 (with a peak of 80); alternatively, it is also possible to exploit wide surveys like SDSS-BOSS \citep{dawson2013}, where the resolution is $R\sim2000$ but the S/N of single galaxies is much lower, and take advantage of the extremely large statistics to significantly increase the S/N stacking spectra of accurately selected passive galaxies.
In this case, the data quality would allow us (i) to carefully select them (excluding galaxies with residual evidence of on-going star formation from the analysis of their spectra), (ii) to precisely determine their physical properties \citep[e.g., in][it was shown with independent approaches that ages and SFHs are accurately recovered without significant systematic offsets from S/N$\gtrsim10$ \AA$^{-1}$]{choi2014,citro2016,carnall2019}, and (iii) to accurately measure their metallicity \citep[one can assume conservatively to a 5\% accuracy; see,  e.g.,][]{moresco2016}.
Therefore, the contribution to the systematic error due to a residual young component can be shown to be negligible \citep{moresco2018}, as well the contribution due to SFH (their SFH would be precisely known). High S/N and spectral resolution spectra would also enable us to perform a comparison between different SPS models, at least to discard the most discordant model (like in the odd-one-out approach) and verify which stellar library better reproduces the data \citep[for an example, see][]{ge2019}. 
In this way, considering the values in Table~\ref{tab:res_mean} and shown in Figures~\ref{fig:offset} and \ref{fig:offset_met}, the systematic uncertainty on $H(z)$ will be :
\begin{equation}
\rm \sigma_{syst}=\pm 0.5\% (IMF) \pm 5.1\% (SPS) \pm 1.9\% (met.)
\end{equation}
yielding a total $\sigma^{tot}_{syst}=5.5$\% if summed in quadrature.

\paragraph{Worst-case scenario} 

Let us assume a lower-quality measurement at the same redshift, where the spectral resolution and S/N would not allow us to do a precise determination of the metallicity (we will consider here an uncertainty of 10\%) or a model selection. In this case, we will also include the contribution due to SFH uncertainty \citep{moresco2016}, and we will use the more modern stellar library. Here we would obtain
\begin{eqnarray}
\rm \sigma_{syst}&=&\pm 2.5\% (SFH) \pm 0.5\% (IMF) \nonumber\\
&& \pm 10.8\% (SPS) \pm 9.8\% (met.)
\end{eqnarray}
for a total of $\sigma^{tot}_{syst}=14.8$\% if summed in quadrature.

\paragraph{Current CC data.}
 
We stress here that the current errors associated with CC data \citep{moresco2012,moresco2015,moresco2016} already consider systematic errors due to SFH and metallicity uncertainties, and that in \citet{moresco2018} it was shown that for these data, the contribution due to a residual contamination of a young population is negligible. In this case, therefore, the remaining sources of systematic uncertainties one would have to take into account are the ones depending on IMF and SPS models, considering that one would want to use the more modern stellar libraries \citep[see, e.g.,][]{ge2019}.

For current $H(z)$ measurements with the CC method, assuming a conservative approach, one would have to add a systematic uncertainty between 13.2\% and 3.9\% (from $z=0.2$ to $1.5$, adding in quadrature the IMF and the SPS contribution), while, discarding the most discordant model, one would have to add a systematic uncertainty between 5.4\% and 2.3\% (from $z=0.2$ to $1.5$).


\section{Conclusions}
\label{sec:conclusions}

In this paper, we have computed and presented the full covariance matrix for systematic uncertainties affecting the CC method. Given the fact that we have addressed in previous analyses the systematic error on $H(z)$ due to an uncertainty on the determination of the SFH of the population \citep{moresco2016} and to a residual contamination of an underlying young component in the CC spectra \citep{moresco2018}, we consider here SSP models, and we focus on determining the impact of adopting different SPS models in terms of assumed IMF, stellar library, and model; moreover, we also estimate the impact of metallicity uncertainties.
We use a large suite of different stellar population models to assess the impact of uncertainties in the stellar physics input.

The main results of this article, summarized in Table~\ref{tab:res_tot}, are the following.

\begin{itemize}
\item The systematic errors induced by the variation of the IMF are small and subdominant, being, on average, $<$0.4\%.
\item The systematic errors due to a variation of stellar library are, instead, larger ($\sim$6.6\%, on average); we note, however, that current model comparisons already highlighted that modern stellar libraries provide better results in reproducing observed data \citep[e.g., see][]{ge2019}. In future works, therefore, one could focus the analysis only on the more modern stellar libraries.
\item The choice of the SPS model dominates the systematic error budget with contributions at the 8.9\% level, on average. However, this value is in many cases driven by a particularly discrepant model, and that by removing it, we can further reduce the error to $\sim$4.5\%.
\item We estimate that an $\sim$ 10\% ($\sim$5\%) error on the determination of the stellar metallicity results in a 9\% (4\%) level error on $H(z)$. The impact on the $H(z)$ determination of an uncertainty on stellar metallicity can, in principle, be kept under control with high spectral resolution and high-S/N data. As an example, \citet{moresco2016} demonstrated  that the metallicity of passively evolving galaxies can be determined at the $\sim 5-10\%$ even when leaving the SFH completely free. 
\item For illustrative purposes, we have finally explored three scenarios of a real measurement that could be performed at $z\sim0.8$ in order to give an example of the potential total systematic error that can be obtained. We found for the best-case scenario a total $\sigma^{tot}_{syst}=5.5$\% and for the worst-case scenario a $\sigma^{tot}_{syst}=14.8$\%. For current data, we show that the additional systematic error to be added to the already considered systematic errors could be at most between 13.2\% and 3.9\% as a function of redshift (in a conservative approach) and between 5.4\% and 2.3\% not considering the outlier model at each redshift.
\end{itemize}
It is worth emphasizing that, in principle, systematic uncertainties can be further reduced with an improvement in SPS modeling. A concerted effort aimed at cross-checking and validating the available models in the literature could result in better and more convergent models. As a consequence, systematic errors on CCs that at the moment are driven by these differences could be minimized.
We also note that the approximation of a piecewise linear $D4000_n-$age relation used to derive Equation~\ref{eq:Hz_D4000} could be further improved once  better convergence of the theoretical models is achieved, exploring at that point different approaches that are less stable at the moment.
We therefore envision that further improvement in SPS modeling might significantly reduce the systematic errors, opening the possibility of obtaining  a percent-level estimate of the expansion rate of the universe over the $0.2 < z < 2$ redshift range with the CC method.

\acknowledgments
{
M.M., A.C. and L.P. acknowledge the grants ASI n.I/023/12/0,  ASI n.2018-23-HH.0, and PRIN MIUR 2015. L.V. acknowledges support by the European Union's Horizon 2020 research and innovation program ERC (BePreSySe, grant agreement 725327). Funding for this work was partially provided by the Spanish MINECO under project PGC2018-098866-B-I00. We thank the anonymous referee for the constructive and useful report that helped to improve the presentation of the results.
}
%

\vspace{5mm}


\software{BC16, BC03 \citep{bruzual2003}, M11 \citep{maraston2011}, E-MILES \citep{vazdekis2016}, FSPS \citep{conroy2009,conroy2010}, pwlf \citep{pwlf}}



\appendix
\section{Extending the Estimate of the Piecewise Linear Slope across the Knees}
\label{sec:slope_btw_knees}

As discussed in Section~\ref{sec:measslope}, a piecewise linear slope is found to be well defined between two knees of the $D4000_n-$age relation. However, some ambiguity might arise across the knees.

\begin{figure}[b!]
	\centering
	\includegraphics[width=0.45\textwidth]{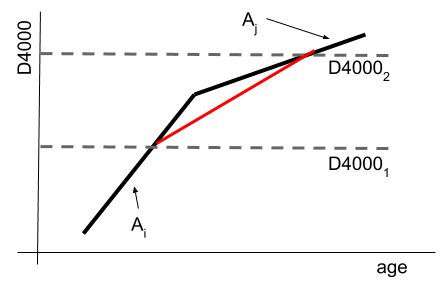}
	\includegraphics[width=0.45\textwidth]{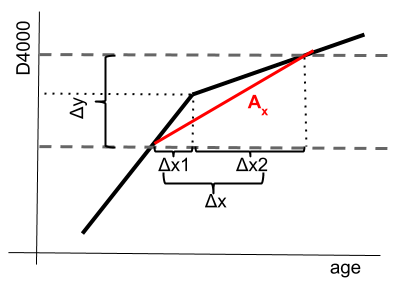}
	\caption{Estimating the correct piecewise linear slope across a knee. In the left panel, a schematic view of the $D4000-$age relation is shown in black, where the dashed lines show the case of a measurement where the couple of $D4000$ values are located across a knee of the relation. In this case, it is incorrect to adopt the slopes $A_i$ or $A_j$ of the lower and upper part of the relation, but another slope should be estimated (shown in red). In the right panel, the various components of Equations~\ref{eq:btwknees1} and \ref{eq:btwknees2} are shown, and as well as the position of the knee (horizontal dashed line).}
	\label{fig:Dnage_intm}
\end{figure}
 
To define $\Delta D4000_n$ in Equation~\ref{eq:Hz_D4000}, two measurements $D4000_1$ and $D4000_2$ are needed. When these two values are all contained in a common range of a piecewise linear fit, the slope is therefore univocally determined. As shown in Figure~\ref{fig:Dnage_intm}, when the couple of $D4000$ values are across a knee, i.e. $D4000_1$ is associated to the slope $A_i$ and $D4000_2$ to the slope $A_j$; a suitable prescrition is needed to determine the correct (and unknown) slope $A_x$.

Considering a mean slope between $A_i$ and $A_j$ would be a poor approximation, since the real slope $A_x$ would depend on the relative position of $D4000_{1}$ and $D4000_{2}$ with respect to the knee. Moreover, we cannot rely on a measurement of the x-axis, since in a real measurement, it is unknown, and we want to express all relevant quantities as a function of the measurable y-axis (the $D4000$).

The unknown slope $A_x$, by definition, can be written as
\begin{equation}
A_{x}=\frac{\Delta y}{\Delta x}=\frac{\Delta y}{\Delta x_1+\Delta x_2}\; ;
\label{eq:btwknees1}
\end{equation}
the two quantities $\Delta x_1$ and $\Delta x_2$ can be expressed as
\begin{eqnarray}
\Delta x_1&=&\frac{A_{i}}{\Delta y_1}\nonumber\\
\Delta x_2&=&\frac{A_{j}}{\Delta y_2}\; .
\label{eq:btwknees2}
\end{eqnarray}
Given that $\Delta y=D4000_{2}-D4000_{1}$, we can rewrite
\begin{eqnarray}
\Delta y_1&=&D4000_{intercept}-D4000_{1}\nonumber\\
\Delta y_2&=&D4000_{2}-D4000_{intercept}\; ,
\label{eq:btwknees3}
\end{eqnarray}
and combining Equations~\ref{eq:btwknees1}, \ref{eq:btwknees2} and \ref{eq:btwknees3} we obtain
\begin{equation}
A_{x}=\frac{\Delta y\cdot A_{i}\cdot A_{j}}{A_{i}\cdot \Delta y_2+A_{j}\cdot \Delta y_1}
\end{equation}

\section{Measurement of the Interpolated Slope}
\label{sec:intpslope}

In this appendix, we explore how the results may change upon a different choice for estimating of the slope of the $D4000_n-$age relations.

We recall that the slope $A$ is defined as the derivative of the $D4000_n$ as a function of age $t$, $d D4000_n/d t$. In our analysis, we fit $D4000_n(t)$ as a piecewise linear function. Here, to obtain a smoother response to the change of the slope as a function of the measured $D4000_n$, we try to fit for $A(D4000_n)$. We choose to fit these relations (rather than the slope as a function of the age) to have  a more direct mapping between the interpolated slope and the measured $D4000$; note that, given this choice, the case of the piecewise linear slope (where we instead fit the $D4000_n-$age relations) is not a subcase of the one presented here.

For this purpose, we first estimate $A$ as a finite difference from the sampled $D4000_n-$age relations, by considering nonadjacent points (the kth and the (k+5)th), to avoid large oscillations in the derivative due to small fluctuations in the relations.

\begin{figure}[b!]
\centering
\includegraphics[width=0.95\textwidth]{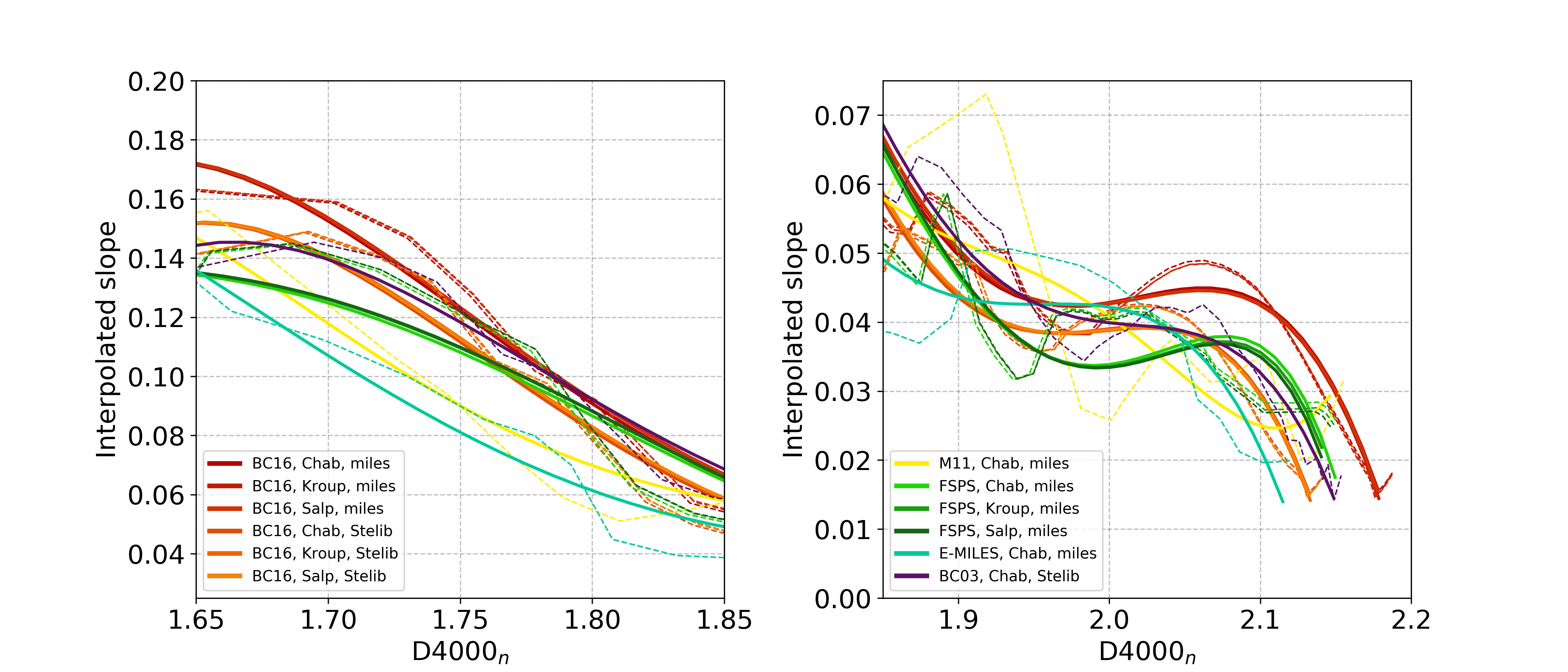}
\caption{Interpolated slopes of the $D4000_n-$age relations shown in Figure~\ref{fig:Dnage2} as a function of $D4000_n$ (dashed lines). The solid lines represent the fifth-order polynomial fit to the relations. The left panel shows a zoom in the range $1.65<D4000_n<1.8$, and the right panel shows a zoom in the range $1.85<D4000_n<2.2$.}
\label{fig:Dnage_slopes}
\end{figure}

We then fit the resulting slope, i.e., $A(D4000_n)$ at a fixed metallicity, with a fifth-order polynomial,
\begin{equation}
{\rm A}(D4000_n)=\sum_{k=0}^{5}a_{k}D4000_{n}^{k}
\label{eq:interp}
\end{equation}
limiting the analysis to the range $D4000_n>1.5$, as this range matches that spanned by current datasets for CCs \citep[see][]{moresco2012,moresco2015,moresco2016}.
We note that the chosen parameterization reproduces well the behavior of the local slope, with a value of $<r^2>=0.965\pm0.019$, with a minimum value of 0.928 and a maximum value of 0.983; the results are shown in Figure~\ref{fig:Dnage_slopes}, and the best-fit parameters are given in Table~\ref{tab:slopes_intp}.

\begin{table*}[t!]
\begin{center}
\caption{Parameters of the Interpolated Slope of the $D4000_n-$Age Relations for Solar Metallicities.}
\begin{tabular}{ccccccc}
\hline
\hline
Model & $a_5$ & $a_4$ & $a_3$ & $a_2$ & $a_1$ & $a_0$ \\
\hline
BC16, miles, chab & 14.70 & -156.59 & 657.62 & -1362.88 & 1395.26 & -564.76\\
BC16, miles, kroup & 13.32 & -143.67 & 609.18 & -1272.54 & 1311.42 & -533.79\\
BC16, miles, salp & 13.01 & -140.35 & 595.56 & -1244.77 & 1283.36 & -522.54\\
BC16, stelib, chab & 20.87 & -219.73 & 913.46 & -1876.93 & 1908.00 & -768.07\\
BC16, stelib, kroup & 18.87 & -201.12 & 844.35 & -1749.10 & 1790.28 & -724.91\\
BC16, stelib, salp & 18.63 & -198.42 & 832.76 & -1724.59 & 1764.71 & -714.36\\
M11, miles, chab & 55.02 & -518.19 & 1946.47 & -3644.18 & 3399.66 & -1263.82\\
FSPS, miles, chab & -38.85 & 351.95 & -1268.92 & 2275.94 & -2031.00 & 721.67\\
FSPS, miles, kroup & -40.65 & 368.83 & -1332.06 & 2393.56 & -2140.12 & 762.00\\
FSPS, miles, salp & -41.64 & 377.85 & -1364.89 & 2453.16 & -2194.08 & 781.49\\
Vazd, emiles, chab & 5.45 & -61.62 & 267.49 & -562.70 & 576.98 & -231.40\\
BC03, stelib, chab & 8.76 & -101.94 & 457.87 & -1001.12 & 1071.22 & -450.21\\
\hline
\end{tabular}
\label{tab:slopes_intp}
\end{center}
\end{table*}

\begin{figure*}
\centering
\includegraphics[width=0.48\textwidth]{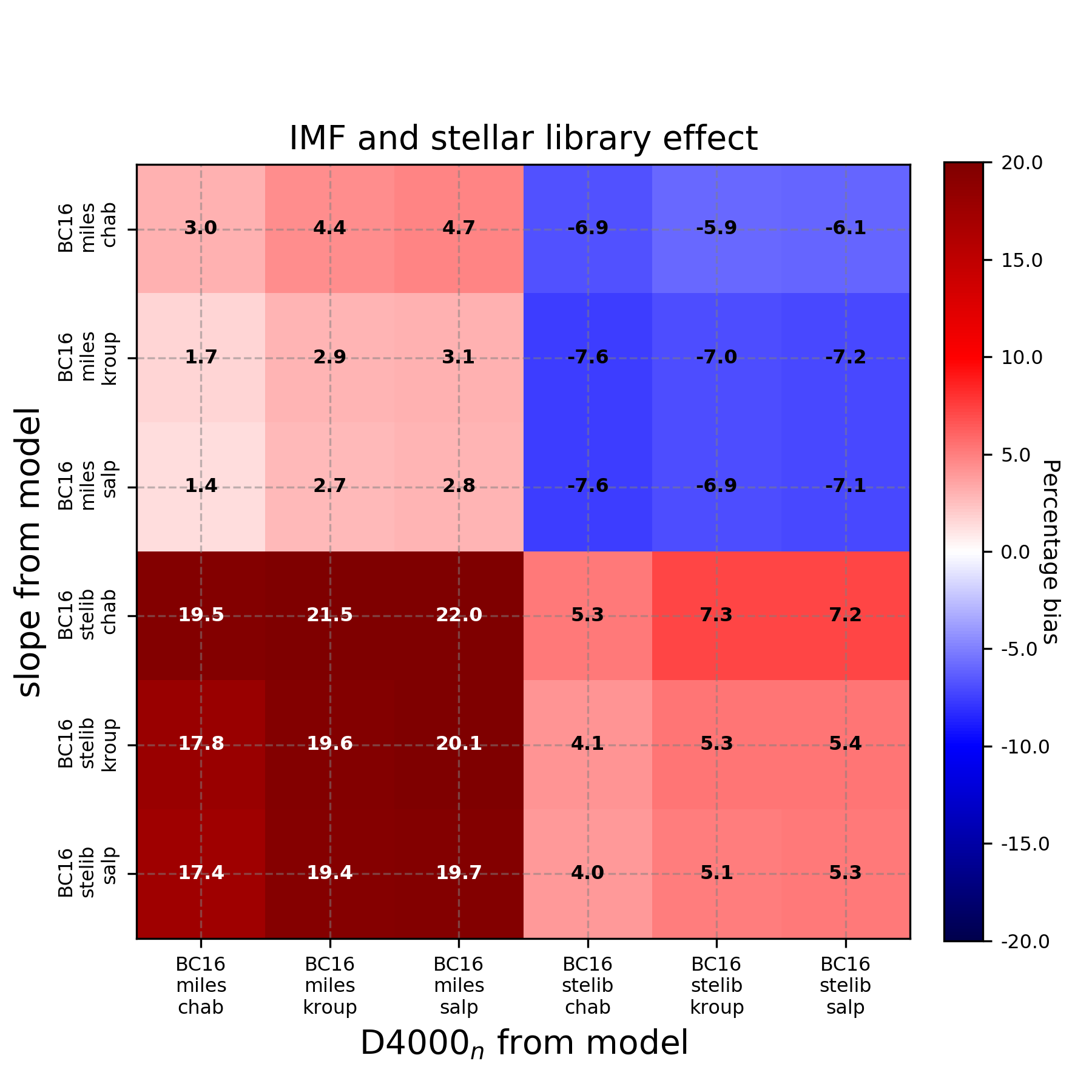}
\includegraphics[width=0.48\textwidth]{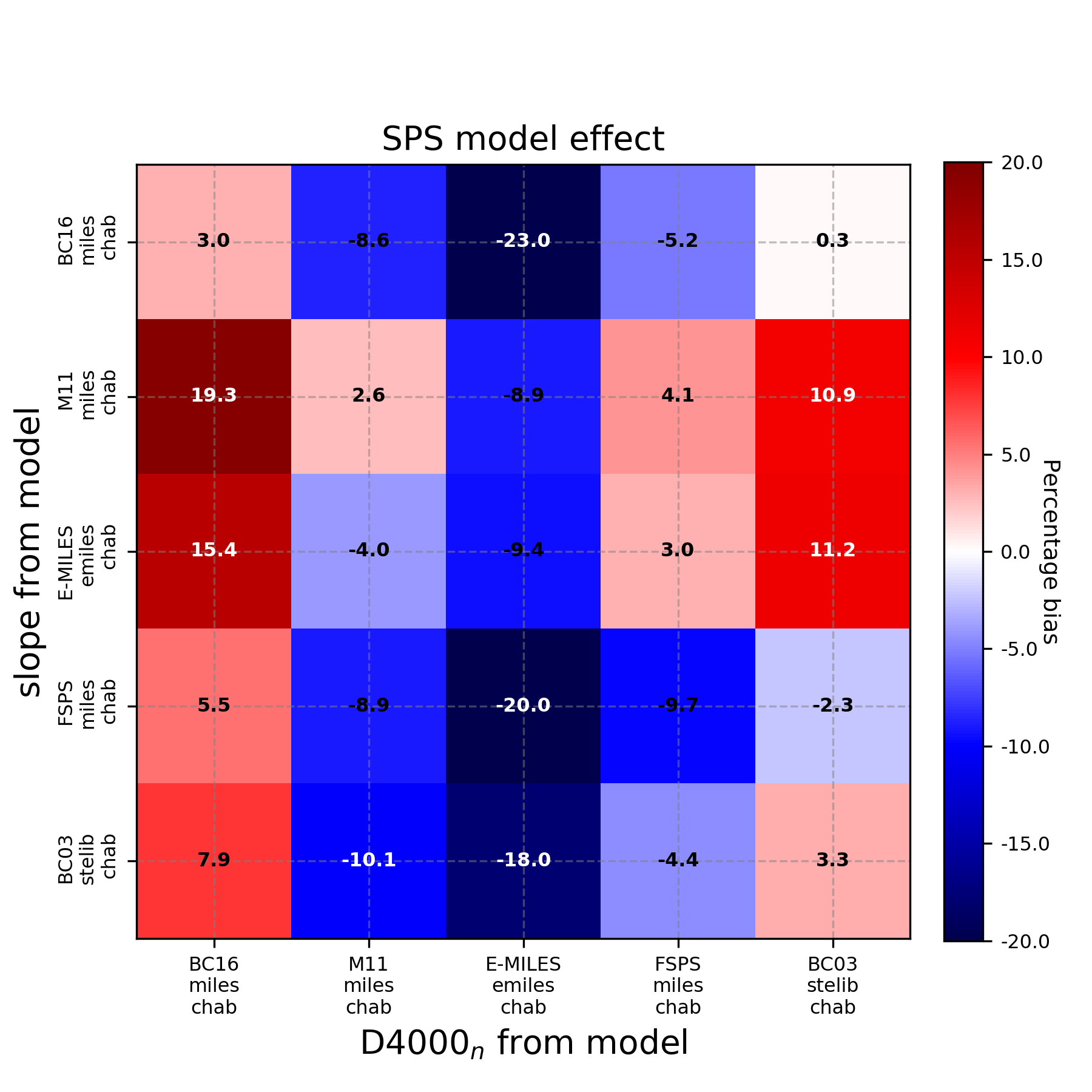}
\caption{Same as Figure~\ref{fig:eta}, but adopting the interpolated slope approach.}
\label{fig:eta_intp}
\end{figure*}

These slopes are then used to estimate the $\eta(z)_{a,b}$ matrices as discussed for the piecewise linear slope in Section~\ref{sec:measmatrix}. The resulting matrices at the same redshift as in Section~\ref{sec:measmatrix} are shown in Figure~\ref{fig:eta_intp}. The errors obtained with this approach are approximately a factor of 2 worse than the ones reported from the piecewise linear slope.
A great part of this variation is due to the fact that in this approach, the slopes obtained with the interpolation have been compared with the raw $D4000_n-$age data, since there was no clear way to invert the relation to obtain a smoothed version of those.
Our interpretation is that the intrinsic noise in the measurement of  $D4000_n$  from the spectra provided by a given  model is comparable to the one induced by  using a different SPS model to measure the $D4000_n$ variations with redshift or age. 

As an alternative, we also studied the possibility of fitting the $A(D4000_n)$ relations with a cubic spline. This approach has the advantage of even more accurately reproducing the original relations, but, as a drawback, it is not possible to provide coefficients to reproduce the relations. With this method, we find slightly better results than with the polynomial fit, but still worse than the piecewise linear fit.

We consider that it is not worth exploiting this approach further, since, unlike the piecewise slope, it is not optimal in the context of CCs. In fact, it is tied to an absolute value of $D4000_n$ to estimate the slope; hence, it is not a purely differential approach, which, as extensively discussed in, e.g., \citet{moresco2012} and \citet{moresco2016}, is one of the most valuable strengths of the method.

\bibliographystyle{aasjournal} 
\bibliography{bib.bib}



\end{document}